\documentclass[11pt]{article}
\usepackage{amsmath,amssymb}

\def\tabaddress#1{{\small\it\begin{tabular}[t]{c}#1
\\[1.2ex]\end{tabular}}}

\font\fr=eufm10 scaled \magstep 1 %(caracteres goticos)
\newtheorem{teor}{Theorem}
\newtheorem{prop}{Proposition}
\newtheorem{corol}{Corollary}

\newtheorem{state}{Statement}
\newtheorem{assum}{Assumption}
\newtheorem{remark}{Remark}
\def\beq{\begin{equation}}
\def\eeq{\end{equation}}
\def\bea{\begin{eqnarray}}
\def\eea{\end{eqnarray}}
\def\beann{\begin{eqnarray*}}
\def\eeann{\end{eqnarray*}}
\def\ben{\begin{enumerate}}
\def\een{\end{enumerate}}
\def\bit{\begin{itemize}}
\def\eit{\end{itemize}}
\def\dst{\(\displaystyle} %\)
\def\derpar#1#2{\frac{\partial{#1}}{\partial{#2}}}

\def\feble#1{\mathrel{\mathop =\limits_{#1}}}

\def\moment#1#2#3{{#1}_{#2}, \ldots, {#1}_{#3}}

\def\qed{\ifvmode\removelastskip\fi
{\unskip\nobreak\hfil\penalty50\hbox{}\nobreak\hfil \hbox{\vrule
height1.2ex width1.2ex}\parfillskip=0pt \finalhyphendemerits=0
\par\smallskip}}
\def\vf{\mbox{\fr X}}
\def\df{{\mit\Omega}}
\def\Lag{{\cal L}}

\def\d{{\rm d}}

\def\Real{\mathbb{R}}

\def\inn{\mathop{i}\nolimits}
\def\Tan{{\rm T}}

\def\ls{(J^1\pi,\Omega_\Lag )}

\def\hso{({\cal P},\Omega_{\rm h}^0)}

\def\Cinfty{{\rm C}^\infty}
\def\proof{( {\sl Proof} )\quad}
\begin{document}

\title{PRE-MULTISYMPLECTIC CONSTRAINT ALGORITHM FOR FIELD THEORIES}
\author{\sc Manuel de Le\'on\thanks{{\bf e}-{\it mail}:
 mdeleon@imaff.cfmacc.csic.es}
 \\
 \tabaddress{Instituto de Matem\'aticas y F\'\i sica Fundamental, CSIC\\
   C/ Serrano 123. E-28006 Madrid. Spain}
   \\
{\sc Jes\'us Mar\'\i n-Solano \thanks{{\bf e}-{\it mail}:
 jmarin@ub.edu}},
 \\
 \tabaddress{Departamento de Matem\'atica Econ\'omica, Financiera
  y Actuarial, UB\\
  Av. Diagonal 690. E-08034 Barcelona. Spain}
   \\
{\sc Juan Carlos Marrero \thanks{{\bf e}-{\it mail}:
 jcmarrer@ull.es}},
 \\
 \tabaddress{Departamento de Matem\'atica Fundamental, Fac. Matem\'aticas,
  U. La Laguna\\
  La Laguna, Tenerife. Spain}
   \\
{\sc Miguel C. Mu\~noz-Lecanda\thanks{{\bf e}-{\it mail}:
 matmcml@ma4.upc.edu}},
{\sc Narciso Rom\'an-Roy\thanks{{\bf e}-{\it mail}:
 matnrr@ma4.upc.edu}},
 \\
 \tabaddress{Departamento de Matem\'atica Aplicada IV\\
  Edificio C-3, Campus Norte UPC\\
  C/ Jordi Girona 1. E-08034 Barcelona. Spain}}
\date{\today}

\pagestyle{myheadings}
\markright{\sc M. de Le\'on {\it et al\/},
   \sl Pre-multisymplectic constraint algorithm for field theories.}
\maketitle
\thispagestyle{empty}

\begin{abstract}
We present a geometric algorithm for obtaining consistent
solutions to systems of partial differential equations,
mainly arising from singular covariant first-order classical field theories.
This algorithm gives an intrinsic description of all the constraint submanifolds.

The field equations are stated geometrically, either representing their
solutions by integrable connections or, what is equivalent, by
certain kinds of integrable $m$-vector fields. First, we consider
the problem of finding connections or  multivector fields
solutions to the field equations in a general framework: a
pre-multisymplectic fibre bundle (which will be identified with
the first-order jet bundle and the multimomentum bundle when
Lagrangian and Hamiltonian field theories are considered). Then,
the problem is stated and solved in a linear context, and a
pointwise application of the results leads to the algorithm for
the general case. In a second step, the integrability of the
solutions is also studied.

Finally, the method is applied to Lagrangian and Hamiltonian field theories and,
for the former, the problem of finding holonomic solutions is also analized.
\end{abstract}


\bigskip

\vbox{\raggedleft AMS s.\,c.\,(2000):
35Q99, 37J05, 53C05, 53D99, 55R10, 58A20, 70S05.
 \\
PACS (1999):
 02.40.Vh, 11.10.Ef.
}\null

 \clearpage

 \tableofcontents

 \section{Introduction}

Systems of singular differential equations have been a matter
of increasing interest, especially during the last 30 years,
and they have been studied separately
in theoretical physics and in some technical areas such as engineering
of electric networks or control theory.
The fundamental characteristic of these kinds of systems is that
the existence and uniqueness of solutions are not assured.

In particular, this situation arises in mechanics when
dynamical systems described by singular Lagrangians are considered.
Furthermore, these systems do not have a nice Hamiltonian description,
since not all the momenta are available, and
there is a submanifold of the momentum phase space
where, in general, the dynamical equations have no solution everywhere.
The same problems arise when considering systems of PDE's associated with field
theories described by singular Lagrangians
(indeed, many field theories are singular, for instance electromagnetism),
as well as in some other applications related with optimal control
theories.

Dirac \cite{Dir-64} was pioneering in solving the problem
for the Hamiltonian formalism of singular mechanical systems,
by developing a constraint
algorithm which gives, in the favourable cases, a final constraint submanifold
where admissible solutions to the dynamics exist
(in the sense that the dynamical evolution remains on this manifold).
 Dirac's main aim was to apply this procedure to field theories.
After Dirac, a lot of work was done in order to geometrize his algorithm.
The first important step was the work by Gotay {\it et al}
\cite{GNH-78}, and its application to the Lagrangian formalism
\cite{GN-79,GN-80}. Other algorithms were given later, in order to find
consistent solutions of the dynamical equations
in the Lagrangian formalism of singular systems
(including the problem of finding holonomic solutions)
\cite{BGPR-86,Ka-82,MR-92}, and
afterwards, new geometric algorithms
were developed to be applied both in the Hamiltonian and
the Lagrangian formalisms \cite{GP-92,HRT-76,
KU-96,MMT-97,Mu-89,SR-83}.

The Lagrangian and Hamiltonian descriptions of field theories,
termed the multisymplectic approach, is the natural extension of
time-dependent mechanics. Therefore, in order to understand
the constraint algorithm for field theories
in a covariant formalism, the first step was
to develop the algorithmic procedures for time-dependent systems.
This work was provided in
\cite{CF-93,CLM-94,GMS-97b,GM-2005,ILDM-99,
LMM-96b,LMMMR-2002,LMM-96c,MS-98,MS-00b,MassVig-00,Vig-00}.
A basic geometric study of these systems
can be found in \cite{Cr-95}.
Furthermore, a qualitative description of constraint algorithms
for field theories was made in \cite{EMR-96,EMR-99b}.

Working within the framework of the multisymplectic description for
these theories, we present in this paper a geometric algorithm for
finding the maximal submanifold where there are consistent
solutions to the field equations of singular theories. This
algorithm gives an intrinsic description of all the constraint
submanifolds. The problem is stated in a generic
pre-multisymplectic fibre bundle, in order to give a solution to both
Lagrangian and Hamiltonian field theories, as well as other
possible kinds of systems of partial differential equations. In
this framework, the solutions to these equations are given
geometrically by integrable connections or, what is equivalent, by
integrable locally decomposable $m$-vector fields which are
transverse to the fibre projection. The key point consists in
using an auxiliar connection for constructing different
geometrical structures needed to develop the algorithm,
by following the same methods introduced in \cite{LMMMR-2002} for
time-dependent singular systems. This technique (the use of a
connection) was used for the first time in \cite{CCI-91}, in order
to obtain (global) Hamiltonian functions, and afterwards applied
both in the Lagrangian and Hamiltonian formalisms for this and
other purposes (see
\cite{EMR-sdtc,EMR-96,GMS-97b,MS-00b,MS-00c,Sd-98}). An exhaustive
use of this technique in mechanics and field theory
can be found in \cite{MS-98,MS-00a,Sd-95}.

First, the problem is reduced to another in the realm of linear algebra,
and solved in this context, and then the results are applied to the
general pre-multisymplectic framework.
In this way,
a constraint algorithm can be developed giving a
sequence of submanifolds which,
in the best case, ends in some final constraint
submanifold where field equations have consistent solutions
(connections or $m$-vector fields), although not necessarily integrable.
The problem of integrability is considered and solved separately.
Finally, Lagrangian and Hamiltonian field theories are
particular cases where the above results are applied straightforwardly,
although in the Lagrangian case the problem of finding holonomic
solutions must be also analized.

The paper is organized as follows:

First, in Section \ref{lt}, we state and solve the algebraic
version of the problem. Then, in Section \ref{gmc},
we pose the general problem
in the context of a pre-multisymplectic fiber bundle and,
applying the results obtained in the previous Section,
the solution is achieved after studying the additional problem
of integrability.
After this, Section \ref{alhft} is devoted to giving
the application to Lagrangian and Hamiltonian field theories,
including the problem of finding holonomic solutions
in the Lagrangian formalism.
Finally, as a classical example, field theories described
by affine Lagrangians are analyzed in Section \ref{ex}.
An Appendix about multivector fields and connections
is also included, in order to make the paper more
self-contained and readable.

 Manifolds are real, paracompact,
 connected and $C^\infty$. Maps are $C^\infty$. Sum over crossed repeated
 indices is understood.

\section{Linear theory}
\protect\label{lt}

\subsection{Statement of the problem. Equivalences}

The problem we want to solve can be first posed and solved
in a linear algebraic way.
In fact, let ${\cal W}$ and ${\cal E}$ be $\Real$-vector spaces
(although, instead of $\Real$, another field of characteristic
different from $2$ can be used), with $\dim\,{\cal E}=m$, and
$\dim\,{\cal W}=m+n$. Let $\sigma\colon{\cal W}\to{\cal E}$ be a
surjective morphism, and denote ${\rm V}(\sigma)=\ker\,\sigma$, and by
$\jmath\colon{\rm V}(\sigma)\hookrightarrow{\cal W}$ the natural
injection. Consider the exact sequence
\beq
0\longrightarrow{\rm V}(\sigma)
\begin{picture}(26,15)(0,0)
\put(12,7){\mbox{$\jmath$}}
\put(3,3){\vector(1,0){20}}
\end{picture}
{\cal W}
\begin{picture}(25,15)(0,0)
\put(12,7){\mbox{$\sigma$}}
\put(3,3){\vector(1,0){20}}
\end{picture}
{\cal E}\longrightarrow 0
\label{exactseq}
\eeq
Suppose that
$\eta\in\Lambda^m{\cal E}^*$ is a volume element; denote
$\omega=\sigma^*\eta$, and assume that a form
$\Omega\in\Lambda^{m+1}{\cal W}^*$ and a subspace ${\cal C}$ of
${\cal W}$ are given. We denote this collection of data as
$(\sigma;\eta,\Omega;{\cal C})$.

Next we consider the following problems in
$(\sigma;\eta,\Omega;{\cal C})$:

\begin{state}
To find a $m$-vector ${\cal X}\in\Lambda^m {\cal C}$ satisfying
that:

1. ${\cal X}$ is decomposable. \qquad
2. $\inn({\cal X})\omega=1$. \qquad
3. $\inn({\cal X})\Omega=0$.
\label{p1}
\end{state}

\begin{state}
To find a subspace ${\cal H}\subset{\cal C}$ satisfying that:

1. $\dim\,{\cal H}=\dim\,{\cal E}=m$. \quad
2. $\sigma\vert_{\cal H}\colon{\cal H}\to{\cal E}$
is an isomorphism. \quad
3. $[\inn(w)\Omega]\vert_{\cal H}=0$,  $\forall w\in{\cal W}$.
\label{p2}
\end{state}

Observe that condition 2 is equivalent to
${\cal W}={\cal H}\oplus{\rm V}\,(\sigma)$.

\begin{state}
To find a linear map ${\bf h}\colon{\cal E}\to{\cal C}\subseteq
{\cal W}$ satisfying that:

1. $\sigma\circ{\bf h}={\rm Id}_{\cal E}$. \qquad
2. $[\inn(w)\Omega]\vert_{{\rm Im}\,{\bf h}}=0$,
for every $w\in{\cal W}$.
\label{p3}
\end{state}

\begin{prop}
Statements \ref{p1}, \ref{p2}, and \ref{p3}
 are equivalent, that is, from
every solution to some of these problems we can obtain
a solution to the others.
\label{equivprob}
\end{prop}
\proof (1 $\Longrightarrow$ 2) \quad Let ${\cal
X}\in\Lambda^m{\cal C}$ be a solution to the problem 1. As a
consequence of the first condition, we have ${\cal
X}=w_1\wedge\ldots\wedge w_m$, with $w_\alpha\in{\cal C}$. If
$e_\alpha=\sigma(w_\alpha)\in{\cal E}$, for every
$\alpha=1,\ldots,m$, by the second condition we have
$\eta(e_1,\ldots,e_m)=1$, and hence $\{e_\alpha\}$ is a basis of
${\cal E}$.

Consider the subspace ${\cal H}=\langle w_1,\ldots,w_m\rangle$.
We have obviously that $\dim\,{\cal H}=m$ and that the restriction
$\sigma\vert_{\cal H}\colon{\cal H}\to{\cal E}$ is an isomorphism.
Furthermore, $[\inn(w)\Omega](w_1,\ldots,w_m)=0$,
for every $w\in{\cal W}$.
Thus ${\cal H}$ is a solution to problem 2.

\quad (2 $\Longrightarrow$ 3) \quad Let ${\cal H}$ be a solution
to problem 2. So $\sigma\vert_{\cal H}$ is an isomorphism. If
$\jmath_{\cal H}\colon{\cal H}\hookrightarrow{\cal C}$ is the
natural injection, let ${\bf h}\colon{\cal E}\to{\cal C}$ be defined as
${\bf h}:=\jmath_{\cal H}\circ(\sigma\vert_{\cal H})^{-1}$.
This map is a solution to problem 3 because
the first condition holds straightforwardly and,
as ${\rm Im}\,{\bf h}={\cal H}$, the second condition holds.

\quad (3 $\Longrightarrow$ 1) \quad Let ${\bf h}$ be a solution to
problem 3. If $\{\moment{e}{1}{m}\}$ is a basis of ${\cal E}$
satisfying that $\eta(e_1,\ldots,e_m)=1$, let $w_\alpha={\bf
h}(e_\alpha)$, and ${\cal X}=w_1\wedge\ldots\wedge w_m$. Then
${\cal X}\in\Lambda^m{\cal C}$ is a solution to problem 1,
because it is decomposable, and
$$
\inn({\cal X})\omega=\omega(w_1,\ldots,w_m)=\eta(e_1,\ldots,e_m)=1.
$$
Furthermore, if $w\in{\cal W}$,
$$
[\inn({\cal X})\Omega](w)=\Omega(w_1,\ldots,w_m,w)=
(-1)^m[\inn(w)\Omega](w_1,\ldots,w_m)=0
$$
since ${\bf h}$ is a solution to problem 3, and $w_{1},
\ldots, w_{m} \in{\rm Im}\,{\bf h}$. \qed

\subsection{Maps induced by a section}
\protect\label{mis}

Consider the exact sequence (\ref{exactseq}),
and let $\nabla\colon{\cal E}\to{\cal W}$
be a section of $\sigma$.
Denote ${\rm H}(\nabla):={\rm Im}\,\nabla$.
We have the splitting
$$
{\cal W}={\rm H}(\nabla)\oplus{\rm V}(\sigma)
$$
${\rm H}(\nabla)$ is called the
{\sl horizontal subspace} of $\nabla$, and ${\rm V}(\sigma)$
is the {\sl vertical subspace} of $\sigma$.
Note that $\sigma\vert_{{\rm H}(\nabla)}$ is an isomorphism.
The above splitting induces the natural projections
$$
\sigma_\nabla^H\colon{\cal W}\to{\rm H}(\nabla)\subset{\cal W} \quad ; \quad
\sigma_\nabla^V\colon{\cal W}\to{\rm V}(\sigma)\subset{\cal W}
$$
with $\sigma_\nabla^H+\sigma_\nabla^V={\rm Id}_{\cal W}$;
and, for every $w\in{\cal W}$, we write
$w=w_\nabla^H+w_\nabla^V$, where $w_\nabla^H\in{\rm H}(\nabla)$
and $w_\nabla^V\in{\rm V}(\sigma)$ are called the
{\sl horizontal} and {\sl vertical components} of $w$
induced by $\nabla$.
In the same way we have the induced splitting
$$
{\cal W}^*={\rm H}^*(\nabla)\oplus{\rm V}^*(\sigma)
$$
where ${\rm H}^*(\nabla)$ is identified with the set
$\{\beta\in{\cal W}^*\ ; \beta\circ\sigma_\nabla^V=0\}$,
and ${\rm V}^*(\sigma)$ with
$\{\beta\in{\cal W}^*\ ; \beta\circ\sigma_\nabla^H=0\}$,
in a natural way.
This splitting of ${\cal W}^*$ induced by $\nabla$
gives rise to a bigradation in $\Lambda^k{\cal W}^*$
given by
$$
\Lambda^k{\cal W}^*=\bigoplus_{p,q=0,\ldots ,k;\ p+q=k}
(\Lambda^p{\rm H}^*(\nabla)\oplus\Lambda^q{\rm V}^*(\sigma))
$$

Now, let ${\cal Z}\in\Lambda^m{\cal E}$ such that $\eta({\cal Z})=1$.
With this condition, ${\cal Z}$ is unique and decomposable,
since $\dim\,{\cal E}=m$. Consider
${\cal Y}^\nabla_\eta=\Lambda^m\nabla({\cal Z})\in\Lambda^m{\cal W}$,
which verifies the following properties:
\begin{enumerate}
\item
${\cal Y}^\nabla_\eta$ is decomposable, because if
${\cal Z}=e_1\wedge\ldots\wedge e_m$, then
${\cal Y}^\nabla_\eta=\nabla(e_1)\wedge\ldots\wedge\nabla(e_m)$.
\item
$\omega({\cal Y}^\nabla_\eta)=1$, since
$$
\omega({\cal Y}^\nabla_\eta)=\sigma^*\eta(\Lambda^m\nabla({\cal Z}))=
\eta[(\Lambda^m\sigma\circ\Lambda^m\nabla)({\cal Z})]=
\eta[\Lambda^m(\sigma\circ\nabla)({\cal Z})]=\eta({\cal Z})=1.
$$
\end{enumerate}
${\cal Y}^\nabla_\eta$ is said to be the
{\sl $m$-vector associated to $\nabla$ and $\eta$},
and it generates $\Lambda^m{\rm H}(\nabla)$.

The bigradation in $\Lambda^k{\cal W}^*$
induces a splitting of $\Omega$ as follows:
$\Omega = \Omega^{(m, 1)} + \Omega^{\nabla}$,
$\Omega^{(m, 1)}$ being a $(m+1)$-form of bidegree $(m,1)$, and
$\Omega^{\nabla}$ a $(m+1)$-form
that includes the rest of components.
Moreover, we have:

\begin{prop}
$\Omega^{(m,1)}=\omega\wedge\gamma^{\nabla}_{\eta}$, where
$\gamma^{\nabla}_{\eta}:=\inn({\cal Y}^{\nabla}_{\eta})\Omega$.
Then $\Omega=\Omega^{\nabla}+\omega\wedge\gamma^{\nabla}_{\eta}$.
\label{gamma}
\end{prop}
\proof
As ${\cal Y}^\nabla_\eta$ generates $\Lambda^m{\rm H}(\nabla)$,
it suffices to prove that $\Omega^{(m, 1)}$ and
$\omega\wedge\gamma^{\nabla}_{\eta}$
coincide when acting on ${\cal Y}^\nabla_\eta\wedge v$,
for every $v\in{\rm V}(\sigma)$. Thus, as
$\gamma^{\nabla}_{\eta}$ vanishes on ${\rm H}(\nabla)$, we obtain
\beann
\Omega^{(m, 1)}({\cal Y}^\nabla_\eta\wedge v)&=&
\Omega({\cal Y}^\nabla_\eta\wedge v)=
 [\inn({\cal Y}^{\nabla}_{\eta})\Omega)](v)=
\gamma^{\nabla}_{\eta}(v)
\\
(\omega\wedge\gamma^{\nabla}_{\eta})({\cal Y}^\nabla_\eta\wedge v)&=&
\omega({\cal Y}^\nabla_\eta)\gamma^{\nabla}_{\eta}(v)= \gamma^{\nabla}_{\eta}(v)
\eeann
\qed

Finally, if ${\bf h}\colon{\cal E}\to{\cal C}$ is a linear map, $\nabla$
induces a splitting
${\bf h}={\bf h}_\nabla^H+{\bf h}_\nabla^V$,
where ${\bf h}_\nabla^H=\sigma_\nabla^H\circ{\bf h}$, and
${\bf h}_\nabla^V=\sigma_\nabla^V\circ{\bf h}$.
Then, we introduce the map (endomorphism of ${\cal W}$)
$$
\widetilde{{\bf h}_\nabla^V}={\bf h}_\nabla^V\circ\sigma=
\sigma_\nabla^V\circ{\bf h}\circ\sigma\colon{\cal W}\to
{\rm V}(\sigma)\subset{\cal W}
$$

\subsection{Characterization of solutions}
\protect\label{chs}

In what follows, we assume that:

\begin{assum}
The $(m+1)$-form $\Omega^{\nabla}$ is of bidegree $(m-1,2)$.
Hence
\beq
\Omega = \Omega^{(m,1)} + \Omega^{(m-1,2)}
\label{splitOm}
\eeq
This is equivalent to demanding that
$\inn(v_1)\inn(v_2)\inn(v_3)\Omega=0$,
for every $v_1,v_2,v_3\in{\rm V}(\sigma)$.
\label{sup1}
\end{assum}

Note that if $U$ and $V$ are real vector spaces of finite
dimension then
$U^{*} \otimes V \cong \{{\bf h}: U \to V \vert {\bf h} \; \; \mbox{is
linear} \}$.
Thus, the auxiliar section $\nabla$ induces the $\Real$-bilinear
map \beq
\begin{array}{ccccc}
\flat_{\Omega }^{\nabla} & \colon & {\cal E}^*\otimes{\cal
C} & \to &
({\cal E}^*\otimes H(\nabla))\times{\rm V}^*(\sigma) \\
& & {\bf h} & \mapsto & ({\bf h}_\nabla^H,
\inn(\inn([\widetilde{{\bf h}_\nabla^V}]^*){\cal Y}^\nabla_\eta))
\Omega\vert_{{\rm V}(\sigma)})
\end{array}
\label{bemol} \eeq where $\inn([\widetilde{{\bf
h}_\nabla^V}]^*){\cal Y}^\nabla_\eta$ is the $m$-vector on ${\cal
W}$ defined as follows: for every $\beta^1,\dots,\beta^m\in{\cal
W}^*$,
$$
(\inn([\widetilde{{\bf h}_\nabla^V}]^*){\cal Y}^\nabla_\eta)
(\beta^1,\dots ,\beta^m) := \sum_{\alpha=1}^m
{\cal Y}^{\nabla}_{\eta}(\beta^1,\dots ,
[\widetilde{{\bf h}^V_{\nabla}}]^{t}(\beta^\alpha),\dots,\beta^m)
$$
Observe that, if
${\cal Y}^\nabla_\eta=
w_1\wedge\ldots\wedge w_m$, with $w_\alpha\in{\cal W}$, then
$$
\inn([\widetilde{{\bf h}_\nabla^V}]^*){\cal Y}^\nabla_\eta=
\sum_{\alpha=1}^m
w_1\wedge\ldots\wedge\widetilde{{\bf h}_\nabla^V}(w_\alpha)
\wedge\ldots\wedge w_m
$$

\begin{teor}
\label{t1}
The necessary and sufficient condition for a linear map
${\bf h}\colon{\cal E}\to{\cal C}$ to be a solution to the problem
posed in Statement \ref{p3} is that
\beq
\flat_{\Omega }^{\nabla}({\bf h})=
(\jmath_{{\rm H}(\nabla)}\circ(\sigma\vert_{{\rm H}(\nabla)})^{-1},
-(\gamma_{\eta}^{\nabla})\vert_{V(\sigma)})
\label{nsc}
\eeq
where $\jmath_{{\rm H}(\nabla)}\colon{\rm H}(\nabla)\to{\cal W}$ denotes
the natural injection, and
$\jmath_{{\rm H}(\nabla)}\circ(\sigma\vert_{{\rm H}(\nabla)})^{-1}$
is the horizontal lift associated with $\nabla$.
\end{teor}
\proof ($\Longrightarrow$)\quad Suppose that the linear map ${\bf
h}\colon{\cal E}\to{\cal C}$ is a solution to the problem posed in
Statement \ref{p3}. Consider the linear map $\varphi\colon {\cal
E} \to {\cal W}$ defined by
$$
\varphi:= {\bf h}- \jmath_{{\rm H}(\nabla)}\circ(\sigma\vert_{{\rm
H}(\nabla)})^{-1} \colon{\cal E}\to{\cal W}.
$$
We have that
$$
\sigma\circ\varphi=\sigma\circ{\bf h}-
\sigma\circ\jmath_{{\rm H}(\nabla)}\circ
(\sigma\vert_{{\rm H}(\nabla)})^{-1}={\rm Id}-{\rm Id}=0
$$
and therefore,
$$
{\bf h}^{H}_{\nabla}=
\jmath_{{\rm H}(\nabla)}\circ(\sigma\vert_{{\rm H}(\nabla)})^{-1}
\quad , \quad
{\bf h}^V_{\nabla} = \varphi
$$
Now, suppose that $\moment{e}{1}{m}\in{\cal E}$
such that $\eta(\moment{e}{1}{m})=1$, and
let $w_\alpha=\nabla(e_\alpha)$, for $\alpha=1,\ldots,m$;
thus ${\cal Y}^\nabla_\eta=w_1\wedge\ldots\wedge w_m$.
We obtain that
$$
{\bf h}(e_\alpha)=
{\bf h}_\nabla^H(e_\alpha)+{\bf h}_\nabla^V(e_\alpha)=
w_\alpha+{\bf h}_\nabla^V(e_\alpha)
$$
As ${\bf h}$ is a solution to the problem,
using the splitting (\ref{splitOm}), for every $v\in{\cal W}$, we have
\beann
0 &=& \Omega({\bf h}(e_1),\dots,{\bf h}(e_m),v)=
\Omega(w_1+{\bf h}_\nabla^V(e_1),\dots,w_m+{\bf h}_\nabla^V(e_m),v)
\\ &=&
\Omega^{(m,1)}(w_1,\dots,w_m,v)+
\sum_{\alpha=1}^m\Omega^{(m-1,2)}
(w_1,\ldots,{\bf h}_\nabla^V(e_\alpha),\ldots,w_m,v)
\\ &=&
\gamma^\nabla_\eta(v)+
(\inn(\inn([\widetilde{{\bf h}_\nabla^V}]^*){\cal Y}^\nabla_\eta)
\Omega)(v)
\eeann
and the result follows.

\qquad ($\Longleftarrow$)\quad
 Suppose that there exists a linear map
${\bf h}\colon{\cal E}\to{\cal C}$ such that (\ref{nsc}) holds;
that is,
\beann {\bf
h}_\nabla^H=\sigma_\nabla^H\circ{\bf h}&=&\jmath_{{\rm
H}(\nabla)}\circ(\sigma\vert_{{\rm H}(\nabla)})^{-1},
\\
(\inn(\inn([\widetilde{{\bf h}_\nabla^V}]^*){\cal Y}^\nabla_\eta)
\Omega)\vert_{{\rm V}(\sigma)}= -\gamma^\nabla_\eta\vert_{{\rm
V}(\sigma)} &=& - \inn({\cal
Y}^\nabla_\eta)\Omega\vert_{V(\sigma)}. \eeann First we prove that
${\bf h}$ is a section of $\sigma$. In fact,
$$
\sigma\circ{\bf h}=\sigma\circ({\bf h}_\nabla^H+{\bf h}_\nabla^V)=
\sigma\circ{\bf h}_\nabla^H=\sigma\circ \jmath_{{\rm
H}(\nabla)}\circ(\sigma\vert_{{\rm H}(\nabla)})^{-1}= {\rm
Id}_{\cal E}.
$$
Furthermore, let $\moment{e}{1}{m}\in{\cal E}$, with
$\eta(\moment{e}{1}{m})=1$, and let
$w_\alpha=\nabla(e_\alpha)$, for $\alpha=1,\ldots,m$.
We have ${\cal Y}^\nabla_\eta=w_1\wedge\ldots\wedge w_m$.
Now
$$
{\bf h}(e_\alpha)=
(\jmath_{{\rm H}(\nabla)}\circ(\sigma\vert_{{\rm H}(\nabla)})^{-1})
(e_\alpha)+{\bf h}_\nabla^V(e_\alpha)=
w_\alpha+{\bf h}_\nabla^V(e_\alpha)
$$
and we must prove that, if $w\in{\cal W}$, then $\Omega({\bf
h}(e_1),\ldots,{\bf h}(e_m),w)=0$. Note that, as ${\bf h}$ is a
section of $\sigma$, it induces a splitting ${\cal W}={\bf h}
({\cal E}) \oplus {\rm V}(\sigma)$, and hence $w=w_{\bf
h}^H+w_{\bf h}^V$, where $w_{\bf h}^H\in{\rm Im}\,{\bf h}$ and
$w_{\bf h}^V\in V(\sigma)$. Then
$$
\Omega({\bf h}(e_1),\dots,{\bf h}(e_m),w_{\bf h}^H)=0,
$$
and it suffices to prove that $\Omega({\bf
h}(e_1),\dots,{\bf h}(e_m),v)=0$, for every $v\in{\rm V}(\sigma)$.
In fact,
\beann \Omega({\bf h}(e_1),\dots,{\bf
h}(e_m),v)&=& \Omega(w_1+{\bf h}_\nabla^V(e_1),\dots,w_m+{\bf
h}_\nabla^V(e_m),v)
\\ &=&
\Omega^{(m,1)}(w_1,\dots,w_m,v)+
\sum_{\alpha=1}^m\Omega^{(m-1,2)}
(w_1,\ldots,{\bf h}_\nabla^V(e_\alpha),\ldots,w_m,v)
\\ &=&
(\inn({\cal Y}^\nabla_\eta)\Omega)(v)+
\sum_{\alpha=1}^m\Omega^{(m-1,2)}
(w_1,\ldots,\widetilde{{\bf h}_\nabla^V}(w_\alpha),\ldots,w_m,v)
\\ &=&
\gamma_{\eta}^{\nabla}(v)+ (\inn(\inn([\widetilde{{\bf
h}_\nabla^V}]^*){\cal Y}^\nabla_\eta) \Omega)(v)= 0
\eeann
\qed

Now, from Theorem \ref{t1}, we deduce that:

\begin{corol}\label{r1a}
A linear map ${\bf h}\colon{\cal E}\rightarrow{\cal C}$ is a solution to
the problem posed in Statement \ref{p3}
if, and only if
$$
{\bf h}^H_{\nabla} = \jmath_{{\rm
H}(\nabla)}\circ(\sigma\vert_{{\rm H}(\nabla)})^{-1} \quad , \quad
[\inn(\inn([\widetilde{{\bf h}^V_{\nabla}}]^{t}){\cal
Y}^{\nabla}_{\eta})\Omega]\vert_{{\rm V}(\sigma)} =
-\gamma^{\nabla}_{\eta}\vert_{{\rm V}(\sigma)}
$$
\end{corol}

Let ${\rm V}(\sigma)^0\subseteq{\cal W}^*$ be the annihilator of
${\rm V}(\sigma)$. It is clear that the vector spaces ${\rm
H}^*(\nabla)$ and ${\rm V}(\sigma)^0$ are isomorphic. The
{\sl orthogonal complement} of ${\cal C}$ with respect to $\Omega$
and $\nabla$ is the subspace $({\cal
C}^{\perp})^{\nabla}_{\Omega}$ of $({\cal E}\otimes{\rm
V}(\sigma)^0)\times{\rm V}(\sigma)$ defined by \beq \label{e5a}
({\cal C}^{\perp})^{\nabla}_{\Omega}:= ({\rm Im}\,\flat_{\Omega
}^{\nabla})^0.
 \eeq
 Then, from Theorem \ref{t1}, we obtain:

\begin{teor}
\label{t2a} There exists a solution to the problem posed in
Statement \ref{p3} if, and only if,
 \beq \label{e60}
 {\bf h}^*
(\jmath_{{\rm H}(\nabla)}\circ(\sigma\vert_{{\rm
H}(\nabla)})^{-1})- \gamma_{\eta}^{\nabla}(Z)=0 \quad,  \quad
\mbox{\rm for every $({\bf h}^*,Z)\in({\cal
C}^{\perp})^{\nabla}_{\Omega}$}.
 \eeq
\end{teor}

Note that, if ${\cal C} = {\cal W}$ and $({\cal
W}^{\perp})^{\nabla}_{\Omega}=\{0\}$, then it is clear that
(\ref{e60}) holds. This is the case in the following Proposition:

\begin{prop}
\label{pr1a} If the $(m+1)$-form
$\Omega^{\nabla}\in\Lambda^{m+1}{\cal W}^*$ given by
 \beq
\label{e6a}
 \Omega^{\nabla} = \Omega - \omega\wedge\gamma^{\nabla}_{\eta}
 \eeq
 is $1$-nondegenerate (that is, the map
$\flat_{\Omega^{\nabla}}\colon{\cal W}\to\Lambda^m{\cal W}^*$,
defined by $\flat_{\Omega^{\nabla}}(v)=\inn(v)\Omega^\nabla$,
for every $v\in{\cal W}$, is
injective), then $({\cal W}^{\perp})^{\nabla}_{\Omega}=\{0\}$.
\end{prop}
\proof
Let $({\bf h}^*,Z)\in({\cal W}^{\perp})^{\nabla}_{\Omega}$.
 From the definitions of $\flat_{\Omega}^{\nabla}$
and $({\cal W}^{\perp})^{\nabla}_{\Omega}$ (eqs. (\ref{bemol}) and
(\ref{e5a})), we obtain that
 \beq \label{e7a}
 {\bf h}^*({\bf
h'}^H_{\nabla})+\inn(Z)\inn(\inn([\widetilde{{\bf
h'}^V_{\nabla}}]^*){\cal Y}^{\nabla}_{\eta})\Omega=0
 \eeq
 for every ${\bf h'}\in Lin({\cal E},{\cal W})$. In particular, this
implies that ${\bf h}^*({\bf h'})=0$,
 for every ${\bf h'}\in{\cal E}^*\otimes H(\nabla)$, and hence
${\bf h}^*=0$. Therefore, using (\ref{e7a}), we deduce that
$$
\inn(Z)\inn(\inn([\widetilde{{\bf h}^V_{\nabla}}]^*) {\cal
Y}^{\nabla}_{\eta})\Omega = 0 \ ,\
\mbox{\rm for every ${\bf h}\in{\cal E}^*\otimes{\cal W}$}.
$$
As a consequence, from (\ref{e6a}) and from assumption \ref{sup1}, it
follows that $\inn(Z)\Omega^{\nabla}=0$
and, since $\Omega^{\nabla}$ is $1$-nondegenerate, we have that
$Z= 0$.
\qed

\section{The general multisymplectic case}
\protect\label{gmc}

 \subsection{Statement of the problem}
 \protect\label{algstate}

 The problem we wish to solve arises from the Lagrangian and
Hamiltonian formalisms in field theories,
although other kinds of systems can also be stated in this way.

The general geometrical setting for these kinds of systems
consists in giving a fibred manifold $\kappa\colon F\to M$ (which
in what follows is assumed to be a fibre bundle), where
$\dim\,M=m>1$ and $\dim\,F=n+m$, and $M$ is an
orientable manifold with volume form $\eta\in\df^m(M)$.
 We denote
$\omega=\kappa^*\eta$. We write $(U;x^\mu,y^j)$,
$\mu=1,\ldots ,m$, $j=1,\ldots ,n$, for local charts of
coordinates in $F$ adapted to the fibred structure, and such that
$\omega=\d x^1\wedge\ldots\wedge\d x^m\equiv\d^mx$. Let
$\Omega\in\df^{m+1}(F)$ be a closed form, and consider the triad
$(F,\Omega,\omega)$. The form $\Omega$ is said to be a
{\sl multisymplectic form} if it is $1$-nondegenerate, that is, if the
map $\flat_{\Omega}: \Tan F \longrightarrow \Lambda^m\Tan^*F$,
defined by $\flat_{\Omega}(v)=\inn(v)\Omega$,
for every $v\in{\cal W}$, is injective. In this case, the system
described by the above triad is called a {\sl multisymplectic system}.
 Otherwise, the form is said to be a
{\sl pre-multisymplectic form}, and the system is
 {\sl pre-multisymplectic}.

The problem is stated as follows:

 \begin{state}
 Given a pre-multisymplectic system $(F,\Omega,\omega)$,
 we want to find a submanifold $\jmath_C\colon C\hookrightarrow F$,
 and a $\kappa$-transverse, locally decomposable and integrable $m$-vector field
${\cal X}_C$ along $C$, in the fibration $\kappa\colon F\to M$, such that
\beq\label{E}
\inn({\cal X}_C(y))\Omega(y) = 0 \quad , \quad  \mbox{\rm for every $y\in C$}.
\eeq

First we obviate the {\sl integrability condition}.
Hence the problem consists in finding
a submanifold $C\hookrightarrow F$
and a locally decomposable $m$-vector field
${\cal X}_C\in\vf^m(F)$ along $C$ such that
 \beq
 \inn({\cal X}_C(y))\omega(y)=1
\quad , \quad
 \inn({\cal X}_C(y))\Omega(y)=0
  \quad ,\quad \mbox{\sl  for every $y\in C$} \; .
 \label{fundeqs}
 \eeq
(Note that the first equation implies that ${\cal X}_C$ is
$\kappa$-transverse).
\label{stat1}
\end{state}

Taking into account Remark \ref{r3} in the Appendix and
Proposition \ref{equivprob}, we have:

\begin{prop}
If $C$ is a submanifold of $F$, then there exists a solution to
 the problem stated in Statement \ref{stat1}
if, and only if, at every point $y\in C$,
there is ${\bf h}_{y} \in \Tan^*_{\kappa(y)}M \otimes \Tan_{y}C
\cong Lin (\Tan_{\kappa(y)}M, T_{y}C)$ such that
\ben
\item
${\bf h}_{y}$ is $\kappa$-transverse (that is, it is a connection
along $C$):
\beq \label{e00}
\Tan_y\kappa\vert_{\Tan_yC} \circ{\bf h}_{y} = Id.
\eeq
\item
For every $(X'_1)_{\kappa(y)}, \dots , (X'_m)_{\kappa(y)} \in
\Tan_{\kappa(y)}M$, and $Y_y \in \Tan_yF$,
\beq \label{e1}
\Omega(y)({\bf h}_{y}((X'_1)_{\kappa(y)}), \dots , {\bf
h}_{y}((X'_m)_{\kappa(y)}), Y_y) = 0.
\eeq
\een
\end{prop}

In order to solve this problem, the use of an arbitrary connection
in the fibration $\kappa\colon F \to M$ is required.
 Thus, let $\nabla$ be a
connection in $\kappa\colon F\to M$, and ${\cal
Y}^{\nabla}_{\eta}$ the corresponding locally decomposable
$m$-vector field on $F$ such that $\inn({\cal
Y}^{\nabla}_{\eta})\omega = 1$. As is well-known (see Appendix
and Section \ref{mis}), the connection $\nabla$ induces a
splitting
$$
 \Lambda^k\Tan^*F=\bigoplus_{p,q=0,\ldots
,k;\ p+q=k} (\Lambda^p{\rm H}^*(\nabla)\oplus\Lambda^q{\rm
V}^*(\kappa))
$$
where ${\rm H}(\nabla) \to F$
is the horizontal subbundle associated with the connection
$\nabla$ and ${\rm V}(\kappa) \to F$ is the vertical subbundle of
the fibration $\kappa: F \to M$. Thus, we have that
\[
\Omega = \Omega^{(m, 1)} + \Omega^{\nabla},
\]
$\Omega^{(m, 1)}$ being a $(m+1)$-form of bidegree $(m,1)$ and
$\Omega^{\nabla}$ a $(m+1)$-form.
Moreover, as a straightforward consequence of Proposition \ref{gamma},
we have that:

\begin{prop}
$\Omega^{(m,1)}=\omega\wedge\gamma^{\nabla}_{\eta}$, where
$\gamma^{\nabla}_{\eta}:=\inn({\cal Y}^{\nabla}_{\eta})\Omega$.
Hence $\Omega=\Omega^{\nabla} + \omega\wedge\gamma^{\nabla}_{\eta}$.
\label{gamma1}
\end{prop}

In what follows, we assume that the following condition holds:

\begin{assum}
The $(m+1)$-form $\Omega^{\nabla}$ is of bidegree $(m-1,2)$.

By Proposition \ref{gamma1}, this is equivalent to demanding that
$$
\inn(Z_1)\inn(Z_2)\inn(Z_3)\Omega=0\ , \mbox{\rm for every
$Z_1,Z_2,Z_3\in\vf^{{\rm V}(\kappa)}(F)$}.
$$
\label{sup2}
\end{assum}

\begin{remark}\label{r1}{\rm
The above assumption is justified because this is the situation in
the Lagrangian and Hamiltonian formalism of field theories
(see Propositions \ref{pb1} and \ref{pb1bbis}).
}\end{remark}

\subsection{Conditions for the existence of solutions on a
submanifold of the total space}
\protect\label{solexist}

Taking into account the above considerations,
the necessary and sufficient condition for
the existence of solutions to the problem posed in the
Statement \ref{fundeqs} arises from the results obtained in
Sections \ref{mis} and \ref{chs}.
The key consists in working
at every point of the manifolds involved in this problem.
Thus, if $y\in C$,
the following identifications can be made:
$$
{\cal E}\equiv\Tan_{\kappa(y)}M \quad ,\quad
{\cal W}\equiv\Tan_yF \quad ,\quad
{\cal C} \equiv\Tan_yC \quad ,\quad
{\rm V}(\sigma)\equiv{\rm V}_y(\kappa)
$$
Then we may consider the $\Real$-linear map
$$
\flat_{\Omega}^{\nabla}(y): \Tan_{\kappa(y)}^{*}M \otimes
\Tan_{y}C \to (\Tan_{\kappa(y)}^{*}M \otimes H_{y}(\nabla)) \times
V_{y}^{*}(\kappa)
$$
defined by \beq \flat_{\Omega}^{\nabla}(y)({\bf h}_{y}) = (({\bf
h}_{y})^{H}_{\nabla}, \inn(\inn([\widetilde{({\bf
h}_{y})^V_{\nabla}}]^{t})({\cal Y}^{\nabla}_{\eta}(y)))
(\Omega(y))\vert_{{\rm V}_{y}(\kappa)}). \label{e0} \eeq
Therefore, Theorem \ref{t1} and Corollary \ref{r1a} lead
 to the following results:
\begin{teor}
\label{t1b}
Let $y\in C$. Then, there exists a linear map
${\bf h}_{y}\in\Tan_{\kappa(y)}^*M \otimes\Tan_{y}C$ 
such that (\ref{e00}) and (\ref{e1})
hold if, and only if,
$$
\flat_{\Omega}^{\nabla}(y)({\bf h}_{y}) = ((\Tan_y\kappa_{{\rm H}(\nabla)})^{-1},
-\gamma_{\eta}^{\nabla}(y)\vert_{{\rm V}_{y}(\kappa)})
$$
where $(\Tan_y\kappa_{{\rm H}(\nabla)})^{-1}\colon
\Tan_{\kappa(y)}M \to {\rm H}_{y}(\nabla)$ is the horizontal lift
at $y$ associated with the connection $\nabla$. (Observe that
$(\Tan_y\kappa_{{\rm H}(\nabla)})^{-1}\in\Tan^*_{\kappa(y)}M
\otimes{\rm H}_{y}(\nabla)$).
\end{teor}

\begin{corol}\label{r1'}
If $y\in C$, and ${\bf h}_{y} \in\Tan_{\kappa(y)}^*M \otimes\Tan_{y}C$, then
(\ref{e00}) and (\ref{e1}) hold if, and only if,
\[
({\bf h}_{y})^{H}_{\nabla} = (\Tan_y\kappa_{{\rm H}(\nabla)})^{-1}, \quad
[\inn(\inn([\widetilde{({\bf h}_{y})^V_{\nabla}}]^{t})
({\cal Y}^{\nabla}_{\eta}(y)))
(\Omega(y))]\vert_{{\rm V}_{y}(\kappa)} =
-(\gamma^{\nabla}_{\eta}(y))\vert_{{\rm V}_{y}(\kappa)}.
\]
\end{corol}

\begin{remark}\label{r2}{\rm
If $y\in C$, let ${\rm V}_{y}(\kappa)^0 \subseteq
\Tan_{y}^{*}{\cal W}$ be the annihilator of the vertical subspace
${\rm V}_{y}(\kappa)$ at the point $y$. Then we have that
\[
[Lin (\Tan_{\kappa(y)}M, {\rm H}_y(\nabla))]^* \cong
\Tan_{\kappa(y)}M \otimes {\rm H}^*_{y}(\nabla) \cong
\Tan_{\kappa(y)}M \otimes{\rm V}_{y}(\kappa)^0 \cong Lin
(\Tan_{\kappa(y)}^*M, {\rm V}_{y}(\kappa)^0).
\]
If ${\bf h}_{y}\colon \Tan_{\kappa(y)}M \to {\rm
H}_{y}(\nabla)$ and  ${\bf h}_{y}^*\colon \Tan_{\kappa(y)}^*M \to {\rm
V}_{y}(\kappa)^0$ are linear maps, $\{(X_1)_{\kappa(y)}, \dots ,
(X_m)_{\kappa(y)}\}$ is a basis of $\Tan_{\kappa(y)}M$ such that
$\{\alpha^1_{\kappa(y)}, \dots ,\alpha^m_{\kappa(y)}\}$ is the
dual basis of $\Tan_{\kappa(y)}^*M$, and
\[
\eta(\kappa(y)) = (\alpha^1)_{\kappa(y)} \wedge  \dots  \wedge
(\alpha^m)_{\kappa(y)}, \quad  {\cal X}_{\eta}(\kappa(y)) =
(X_1)_{\kappa(y)} \wedge  \dots  \wedge (X_m)_{\kappa(y)},
\]
then, taking
$(Y_{i})_{y} = (\Tan_y\kappa_{{\rm H}(\nabla)})((X_{i})_{\kappa(y)})$
and $\beta^{i}_{y} =\alpha^{i}_{\kappa(y)} \circ\Tan_y\kappa$,
 for all $i \in \{1, \dots , m\}$,
we deduce that
$\{(Y_1)_{y}, \dots , (Y_m)_{y}\}$
and $\{\beta^1_{y}, \dots , \beta^m_{y}\}$ are a basis
of ${\rm H}_{y}(\nabla)$ and ${\rm V}_{y}(\kappa)^0$, respectively. Moreover, if
\[
{\bf h}_{y}((X_{i})_{\kappa(y)}) =
({\bf h}_{y})^{j}_{i}(Y_{j})_{y}, \quad {\bf h}_{y}^*(\alpha^{i}_{\kappa(y)}) =
({\bf h}_{y}^*)^{i}_{j}\beta^j_y, \quad
\mbox{\rm for every $i \in \{1, \dots , m\}$}
\]
 it follows that $<{\bf h}_{y}^{*}, {\bf h}_{y}> = {\bf h}_{y}^{*}({\bf h}_{y}) =
({\bf h}_{y})^{j}_{i}({\bf h}_{y}^*)^{i}_{j}$.
}\end{remark}

Now, if $y\in C$, the orthogonal complement
$(\Tan_{y}^{\perp}C)^{\nabla}_{\Omega}$ with respect
to $\Omega$ and $\nabla$ is the subspace  of
$(\Tan_{\kappa(y)}M \otimes{\rm V}_{y}(\kappa)^0)\times{\rm V}_{y}(\kappa)$
defined by
\beq
\label{e5'}
(\Tan_{y}^{\perp}C)^{\nabla}_{\Omega} =({\rm Im}\,\flat_{\Omega}^{\nabla}(y))^0.
\eeq

 As in Theorem \ref{t2a},
 from Theorem \ref{t1b} we obtain

\begin{teor}
\label{t2}
Let $y\in C$. Then, there exists a linear map
${\bf h}_{y} \in\Tan_{\kappa(y)}^*M \otimes\Tan_{y}C$ such that
\[
({\bf h}_{y})^{H}_{\nabla} = (\Tan_y\kappa_{{\rm H}(\nabla)})^{-1}, \quad
[\inn(\inn([\widetilde{({\bf h}_{y})^V_{\nabla}}]^{t})
({\cal Y}^{\nabla}_{\eta}(y)))
(\Omega(y))]\vert_{{\rm V}_{y}(\kappa)} =
-(\gamma^{\nabla}_{\eta}(y))\vert_{{\rm V}_{y}(\kappa)}
\]
if, and only if,
\beq
\label{e6}
{\bf h}_{y}^*(\Tan_y\kappa_{{\rm H}(\nabla)})^{-1} -
\gamma_{\eta}^{\nabla}(y)(Z_{y}) = 0,  \quad \mbox{\rm for every} \; \;
({\bf h}_{y}^{*}, Z_{y}) \in (\Tan_{y}^{\perp}C)^{\nabla}_{\Omega}.
\eeq
\end{teor}

Note that if $(\Tan_{y}^{\perp}C)^{\nabla}_{\Omega} = \{0\}$ then
it is clear that (\ref{e6}) holds. Thus, from Proposition
\ref{pr1a}, we have:

\begin{prop}
\label{pr1} If the $(m+1)$-form $\Omega^{\nabla}$ on $F$ given by
$\Omega^{\nabla} = \Omega - \omega \wedge \gamma^{\nabla}_{\eta}$
is $1$-nondegenerate, that is, the map $\flat_{\Omega^{\nabla}}
\colon \Tan F \to \Lambda^{m}\Tan^*F$ is injective, then
\[
(\Tan_{y}^{\perp}F)^{\nabla}_{\Omega} = \{0\}, \quad \quad
\mbox{\rm for every} \; \; y \in F.
\]
\end{prop}

\subsection{The pre-multisymplectic constraint algorithm}
 \protect\label{ca}

 Now we apply the above results in order to solve the problem
 stated in Section \ref{algstate}. The procedure is algorithmic, and
gives a sequence of  subsets $\{ C_i\}$ of $F$. Then, we assume that:

\begin{assum}
 Every subset $C_i$ of this sequence is a regular submanifold of $F$,
 and its natural injection is an embedding.
 \label{asub}
\end{assum}

Thus, we consider the submanifold $C_1\hookrightarrow F$ where a
solution exists, that is,
\[
\begin{array}{lcr}
 C_1 &=& \{ y\in F\ \mid \ \exists {\bf h}_{y}\in Lin
(\Tan_{\kappa(y)}M, \Tan_{y}F) \; \mbox{such that} \; ({\bf
h}_{y})^{H}_{\nabla} = (\Tan_y\kappa_{{\rm H}(\nabla)})^{-1}, \\&&
[\inn(\inn([\widetilde{({\bf h}_{y})^V_{\nabla}}]^{*})({\cal
Y}^{\nabla}_{\eta}(y))) (\Omega(y))]\vert_{{\rm V}_{y}(\kappa)} =
-(\gamma^{\nabla}_{\eta}(y))\vert_{{\rm V}_{y}(\kappa)}\} \; .
\end{array}
\]
Then, using the results of Section \ref{solexist}, we deduce that
there is a locally decomposable section ${\cal X}_{1}$ of the
vector bundle $\Lambda^{m}\Tan_{C_{1}}F \to C_{1}$ such that
$(\inn({\cal X}_{1})\omega)\vert_{C_{1}} = 1$ and $(\inn({\cal
X}_{1})\Omega)\vert_{C_{1}} = 0$. However, in general, ${\bf
h}_{y}(\Tan_{\kappa(y)}M)$ is not a subspace of $\Tan_{y}C_{1}$
and then ${\cal X}_{1}$ is not tangent to $C_{1}$ or, in other
words, in general, ${\cal X}_{1}$ is not a connection in the
fibration $\kappa\colon F \to M$ along $C_{1}$. Therefore, we
consider the submanifold
\[
\begin{array}{lcr}
 C_2 &=& \{ y_{1}\in C_{1}\ \mid \ \exists {\bf h}_{y_{1}}\in Lin
(\Tan_{\kappa(y_{1})}M, \Tan_{y_{1}}C_{1}) \; \mbox{such that} \;
({\bf h}_{y_{1}})^{H}_{\nabla} = (\Tan_{y_1}\kappa_{{\rm
H}(\nabla)})^{-1}, \\&& [\inn(\inn([\widetilde{({\bf
h}_{y_{1}})^V_{\nabla}}]^{*})({\cal Y}^{\nabla}_{\eta}(y_{1})))
(\Omega(y_{1}))]\vert_{{\rm V}_{y_{1}}(\kappa)} =
-(\gamma^{\nabla}_{\eta}(y_{1}))\vert_{{\rm V}_{y_{1}}(\kappa)}\}
\; .
\end{array}
\]
Then, there is a locally decomposable section ${\cal X}_{2}$
of the vector bundle $\Lambda^{m}\Tan_{C_{2}}C_{1} \to C_{2}$ such
that $(\inn({\cal X}_{2})\omega)\vert_{C_{2}} = 1$ and
$(\inn({\cal X}_{2})\Omega)\vert_{C_{2}} = 0$. However, in general,
 ${\cal X}_2$ is not a connection in the fibration
$\kappa\colon F \to M$ along $C_{2}$. Following this process, we
obtain a sequence of constraint submanifolds
 \beq
 \cdots \stackrel{j^i_{i+1}}{\hookrightarrow} C_i
 \stackrel{j^{i-1}_i}{\hookrightarrow} \cdots
 \stackrel{j^1_2}{\hookrightarrow} C_1
 \stackrel{j_1}{\hookrightarrow} C_0\equiv F\; .
 \label{seqsubman0}
 \eeq
 For every $i\geq 1$, $C_i$ is called the {\sl $i$th constraint submanifold}.

 This procedure is called the {\sl pe-multisymplectic constraint
algorithm}. We have two possibilities:
 \bit
 \item
 There exists an integer $k>0$ such that $\dim C_k \leq m-1$. This
 means that the equations have no solution on a submanifold of $F$.
 \item
 There exists an integer $k>0$ such that $C_{k+1}=C_k\equiv C_f$.
In such a case, there exists a connection ${\cal
X}_{f}$ in the fibration $\kappa \colon F \to M$ along $C_{f}$ such that
$$
\inn({\cal X}_{f}(y_{f}))(\Omega(y_{f})) = 0, \quad \mbox{for every} \;
\; y_{f} \in C_{f}.
$$
 In this case, $C_f$ is called the  {\sl final constraint submanifold}.
 This is the situation which is interesting to us. Note that the
existence of a connection in the fibration $\kappa \colon F \to M$
along $C_{f}$ implies that $\kappa(C_{f})$ is an open subset of $M$
and that $\kappa\vert_{C_{f}} \colon C_{f} \to \kappa(C_{f})$ is a
fibration (see Remark \ref{r3} in the Appendix). In particular,
$\dim C_{f} \geq m$.
\eit
 Next we give an intrinsic characterization of the constraints
 which define the constraint submanifolds $C_i$.
For this purpose, we consider the vector bundle over $F$,
\[
W(\kappa, \nabla) = (\kappa^{*}(\Tan^{*}M)\otimes{\rm
H}(\nabla)) \oplus_{F} {\rm V}^*(\kappa)
\]
whose fiber over the point $y \in F$ is
\[
W_{y}(\kappa, \nabla) = (\Tan^{*}_{\kappa(y)}M \otimes{\rm
H}_{y}(\nabla))\times {\rm V}_y^*(\kappa) \cong Lin
(\Tan_{\kappa(y)}M, {\rm H}_{y}(\nabla)) \times {\rm
V}_y^*(\kappa).
\]
The horizontal lift associated with the connection $\nabla$ and
the $1$-form $\gamma_{\eta}^{\nabla}$ induce a section
$((\Tan\kappa_{{\rm H}(\nabla)})^{-1},\linebreak
-(\gamma^{\nabla}_{\eta})\vert_{{\rm V}(\kappa)})$ of this vector
bundle given by
\[
((\Tan\kappa_{{\rm H}(\nabla)})^{-1},
-(\gamma^{\nabla}_{\eta})\vert_{{\rm V}(\kappa)})(y)
= ((\Tan_y\kappa_{{\rm H}(\nabla)})^{-1},
-(\gamma_{\eta}^{\nabla}(y))\vert_{{\rm V}_{y}(\kappa)}),
\quad \mbox{for every} \; \; y \in F.
\]
Furthermore, let $W_{C_{i}}(\kappa, \nabla)$ be the vector
bundle over the submanifold $C_{i}$ whose fiber at the point $y_{i}
\in C_{i}$ is $W_{y_{i}}(\kappa, \nabla)$. Moreover, we may consider
the orthogonal complement
$(\Tan_{y_{i}}^{\perp}C_{i})^{\nabla}_{\Omega}$ of
$\Tan_{y_{i}}C_{i}$ with respect to $\Omega$ and $\nabla$ given by
(see (\ref{e5'}))
\beann
(\Tan_{y_{i}}^{\perp}C_{i})^{\nabla}_{\Omega} = \{({\bf
h}_{y_{i}}^{*}, Z_{y_{i}}) \in Lin (\Tan_{\kappa(y_{i})}^{*}M, {\rm
V}_{y_{i}}(\kappa)^0) \times {\rm V}_{y_{i}}(\kappa) \, \mid \,
{\bf h}_{y_{i}}^{*}(({\bf h}_{y_{i}})^{H}_{\nabla}) +
\\ \inn(Z_{y_{i}})
\inn(\inn([\widetilde{({\bf h}_{y_{i}})^V_{\nabla}}]^{t})({\cal
Y}^{\nabla}_{\eta}(y))) (\Omega(y_{i})) = 0\ , \ \mbox{for every}
\; {\bf h}_{y_{i}} \in Lin (\Tan_{\kappa(y_{i})}M,
\Tan_{y_{i}}C_{i}) \}. \eeann Note that
$(\Tan_{y_{i}}^{\perp}C_{i})^{\nabla}_{\Omega} \subseteq
W_{y_{i}}^{*}(\kappa, \nabla)$. Furthermore, if
$(\Tan^{\perp}C_{i})^{\nabla}_{\Omega}$ is a vector subbundle of
rank $r$ of $W_{C_{i}}^{*}(\kappa, \nabla)$ (that is, the
dimension of $(\Tan^{\perp}_{y_{i}}C_{i})^{\nabla}_{\Omega}$ is
$r$, for every $y_{i} \in C_{i}$) then one may choose a set of $r$
local sections $\{({\bf h}_{1}^{*}, Z_{1}), \dots , ({\bf
h}_{r}^{*}, Z_{r})\}$ of the vector bundle $W^{*}(\kappa, \nabla)
\to F$ such that $\{({\bf h}_{1}^{*}, Z_{1})\vert_{C_{i}}, \dots ,
({\bf h}_{r}^{*}, Z_{r})\vert_{C_{i}}\}$ is a local basis of the
space $\Gamma((\Tan^{\perp}C_{i})^{\nabla}_{\Omega})$ of sections
of the vector subbundle $(\Tan^{\perp}C_{i})^{\nabla}_{\Omega} \to
C_{i}$. In addition, using Theorem \ref{t2}, we deduce
\begin{teor}
\label{t3}
Every submanifold $C_{i}$ $(i \geq 1)$ in the sequence
(\ref{seqsubman0}) may be defined as
\[
 C_i=\{ y_{i-1} \in C_{i-1}\ \mid \ \langle ((\Tan\kappa_{{\rm H}(\nabla)})^{-1},
-(\gamma^{\nabla}_{\eta})\vert_{{\rm V}(\kappa)})(y_{i-1}),
(\Tan^{\perp}_{y_{i-1}}C_{i-1})^{\nabla}_{\Omega}\rangle=0\}.
\]
 Therefore, if $(\Tan^\perp C_{i-1})^{\nabla}_{\Omega}$ is a vector
subbundle of rank $r$ of $W^*_{C_{i-1}}(\kappa, \nabla)$ and
$\{({\bf h}_{1}^{*}, Z_{1})^{(i-1)}, \dots , \linebreak ({\bf
h}_{r}^{*}, Z_{r})^{(i-1)}\}$  is a set of sections of the vector
bundle $W^*(\kappa, \nabla) \to F$ spanning locally the space
$\Gamma((\Tan^{\perp}C_{i-1})^{\nabla}_{\Omega})$, then $C_{i}$,
is defined locally, as a submanifold of $C_{i-1}$,  as the zero set
of the functions $\xi^{(i)}_{j} \in C^{\infty}(F)$ given by
\[
\xi^{(i)}_{j} = ((\Tan \kappa_{{\rm H}(\nabla)})^{-1},
-(\gamma_{\eta}^{\nabla})\vert_{{\rm V}(\kappa)})(({\bf
h}_{j}^{*}, Z_{j})^{(i-1)}).
\]
These functions are called {\rm $i$th-generation constraints}.
\end{teor}

\subsection{The integrability algorithm}
\protect\label{intalg}

Suppose that after applying the premultisymplectic constraint algorithm
we have a final constraint submanifold $C_f\hookrightarrow F$ and a
connection defined by the multivector field
${\cal X}_{f}$ in the fibration $\kappa \colon F \to M$
along $C_{f}$ such that (\ref{E}) holds on $C_{f}$, that is,
\beq
\inn({\cal X}_{f}(y))\Omega(y) = 0, \quad \mbox{for every} \; \; y \in C_{f}.
\label{form1}
\eeq
However, ${\cal X}_{f}$ is not, in general, a flat connection.
Nevertheless, in many cases, one may find a submanifold ${\cal
I}_{f}$ of $C_{f}$ such that $({\cal X}_{f})\vert_{{\cal I}_{f}}$ is a
flat connection in the fibration $\kappa \colon F \to M$ along ${\cal
I}_{f}$ and (\ref{E}) holds for $({\cal X}_{f})\vert_{{\cal I}_{f}}$.

Next we present an algorithm which enables us to find this submanifold
(which is an adapted version of that given in \cite{EMR-98}).
This is a local algorithm, that is, we are in fact working on suitable open sets
in $C_f$. Hence, let ${\cal X}_{f}\equiv\bigwedge_{\mu=1}^m X_\mu$
be a solution to (\ref{form1}).
\bit
\item
{\sl Integrability condition}:
The condition that ${\cal X}_{f}$ is flat is equivalent to
demanding that the distribution spanned by $\moment{X}{1}{m}$ is
involutive. Then, if $c_{f} = dim\,C_{f}$, let
$\moment{Z}{1}{n-m}\in\vf (F)$, such that
$\{\moment{X}{1}{m}, \moment{Z}{1}{c_{f}-m}\}$
is a local basis of the module of vector fields on $C_f$. Therefore,
for every pair $X_\mu,X_\nu$ ($1\leq\mu,\nu\leq m$) we have
\[
[X_\mu,X_\nu]=f_{\mu\nu}^\rho X_\rho+\zeta_{\mu\nu}^lZ_l
\]
for some functions $f_{\mu\nu}^\rho,\zeta_{\mu\nu}^l$.
Consider the system $\zeta^l_{\mu\nu}=0$ and let
\[
{\cal I}_1=\{ y\in C_f\ ;\  \zeta_{\mu\nu}^l(y)=0\ ,\ \forall\mu,\nu,l\}.
\]
We have three options:
\ben
\item
${\cal I}_1=C_f$.
Then the distribution spanned by $\moment{X}{1}{m}$ is involutive,
and $({\cal X}_{f})\vert_{C_{f}}$ is a flat connection in the
fibration $\kappa \colon F \to M$ along $C_{f}$.
\item
${\cal I}_1=\emptyset$.
Then the distribution spanned by $\moment{X}{1}{m}$ is not involutive
at any point in $C_{f}$, and hence the $m$-vector field ${\cal X}_{f}$ is
not integrable.
\item
{\sl ${\cal I}_1$ is a proper subset of $C_f$}.
In this case we assume that ${\cal I}_1$ is a closed submanifold of $C_f$
and the functions $\zeta_{\mu\nu}^l$
are the constraints locally defining ${\cal I}_1$.
The distribution spanned by $\moment{X}{1}{m}$ is involutive on ${\cal I}_1$;
that is, the $m$-vector field ${\cal X}_{f}$ is integrable on ${\cal I}_1$.
\een
If ${\cal X}_{f}$ is tangent to ${\cal
I}_{1}$, then $({\cal X}_{f})\vert_{{\cal I}_{1}}$ defines a flat
connection in $\kappa \colon F \to M$ along ${\cal I}_{1}$ and
(\ref{E}) holds on ${\cal I}_{1}$ which implies that the problem is
solved. Nevertheless, this is not the case in general,
so we need the following:
\item
{\sl Tangency condition}:
Consider the set
\[
{\cal I}_2:=\{ y\in {\cal I}_1 \ ;\ {\cal
X}_{f}(y)\in\Lambda^m\Tan_{y}{\cal I}_1\}
\]
For ${\cal I}_2$ we have the same problem, so we define inductively, for $i>1$,
\[
{\cal I}_i:=\{ y\in {\cal I}_{i-1} \ ;\ {\cal
X}_{f}(y)\in\Lambda^m\Tan_{y}{\cal I}_{i-1}\}
\]
and assume that we obtain a sequence $\ldots \subset {\cal
I}_i\subset\ldots \subset {\cal I}_1\subset C_f$ such that ${\cal
I}_{i}$ is a non-empty (closed) submanifold of $F$, for all $i$, or
${\cal I}_{i} = \emptyset$, for some $i$.

Observe that the locally decomposable $m$-vector field
${\cal X}_{f}=X_1\wedge\ldots\wedge X_m$ is tangent to ${\cal I}_i$
(with ${\cal I}_{i} \neq \emptyset$) if, and only if, $X_\mu$ is tangent to
${\cal I}_i$, for every $\mu$.

Thus, using the constraints, we have that, if
$\{\zeta_{\alpha_i}^{(i)}\}$ is a basis of constraints
defining locally ${\cal I}_i$ in ${\cal I}_{i-1}$, the tangency condition is
$0 \feble{{\cal I}_i} X_\mu(\zeta^{(i)}_{\alpha_i})$
(for every $\mu,\alpha_i$), that is, we have
\[
{\cal I}_{i+1}:=\{ y\in {\cal I}_i \ ;\
X_\mu(\zeta^{(i)}_{\alpha_i})(y)=0 \ ,\ \forall \mu,\alpha_i\}, \quad
\mbox{for every} \; \; i \geq 1.
\]
\eit

The above algorithm ends at step $f$ in one of the following two options:
\ben
\item
$\dim {\cal I}_{f} \leq m-1$. In such a case, we deduce that it is not possible
to find a submanifold ${\cal I}$ of $C_{f}$ such that $({\cal
X}_{f})\vert_{{\cal I}}$ is a flat connection in the fibration $\kappa
\colon F \to M$ along ${\cal I}$. Therefore, we must consider (if it
exists) another connection ${\cal X}'_{f}$ along $C_{f}$ such that
$\inn({\cal X}'_{f}(y))\Omega(y) = 0$, for every $y \in C_{f}$, and then
we must repeat the above procedure.
\item
${\cal I}_{f+1} = {\cal I}_{f}$. In this case
 ${\cal I}_{f}$ is a submanifold of $F$ and we deduce that $({\cal
X}_{f})\vert_{{\cal I}_{f}}$ is a flat connection in the fibration
$\kappa \colon F \to M$ along ${\cal I}_{f}$ such that $\inn({\cal
X}_{f}(y))(\Omega(y)) = 0$, for every $y \in {\cal I}_{f}$. Thus, the
problem is solved. As in Section \ref{ca}, we remark that the existence
of a connection in the fibration $\kappa \colon F \to M$ along ${\cal
I}_{f}$ implies that $\kappa({\cal I}_{f})$ is an open subset of $M$
and that $\kappa\vert_{{\cal I}_{f}} \colon {\cal I}_{f} \to \kappa({\cal
I}_{f})$ is a fibration. In particular, $dim {\cal I}_{f} \geq m$.
\een

We will call this procedure the {\sl integrability algorithm} for
decomposable $m$-vector fields.

\section{Application to Lagrangian and Hamiltonian field theories}
\protect\label{alhft}

\subsection{Lagrangian and Hamiltonian field theories}
\protect\label{Lft}

(For details on the construction of the Lagrangian and Hamiltonian
formalisms of field theories, see for instance, \cite{BSF-88},
\cite{CCI-91}, \cite{EMR-96},
 \cite{EMR-98}, \cite{EMR-99b}, \cite{EMR-00},
 \cite{GMS-97}, \cite{HK-01}, \cite{Ka-98},
 \cite{LMM-96}, \cite{Sd-95}, \cite{Sa-89}.)

A {\sl first-order classical field theory} is described by its {\sl
configuration fibre bundle} $\pi\colon E\to M$ and a {\sl
Lagrangian density} which is a $\bar\pi^1$-semibasic $m$-form, $\Lag$, on
$J^1\pi$ (the first-order jet bundle of $\pi\colon E\to M$).
$\Lag$ is usually written as
$\Lag =L(\bar\pi^{1*}\eta)\equiv L\omega$,
where $L \in\Cinfty (J^1\pi)$ is
the {\sl Lagrangian function} associated with $\Lag$ and $\omega$,
and $\pi^1\colon J^1\pi\to E$ and
$\bar\pi^1:=\pi\circ\bar\pi\colon J^1\pi\to M$
are the natural projections.
The {\sl Poincar\'e-Cartan $m$ and $(m+1)$-forms} associated with
the Lagrangian density $\Lag$ are defined using the {\sl vertical
endomorphism} ${\cal V}$ of the bundle $J^1\pi$
$$
\Theta_{\Lag}:=\inn({\cal V})\Lag+\Lag\in\df^{m}(J^1\pi)
\quad ;\quad
\Omega_{\Lag}:= -\d\Theta_{\Lag}\in\df^{m+1}(J^1\pi)
$$
Then a {\sl Lagrangian system} is a couple $\ls$.
The Lagrangian system is {\sl regular} if
 $\Omega_{\Lag}$ is $1$-nondegenerate.
Elsewhere it is called {\sl singular}.
In a natural chart of coordinates $(x^\alpha,y^A,v^A_\alpha)$ in $J^1\pi$
(adapted to the bundle structure, and such that
$\omega=\d x^1\wedge\ldots\wedge\d x^m\equiv\d x^m$)
we have
\bea
\Omega_{\Lag}&=& \displaystyle -\frac{\partial^2L}{\partial
v^B_\nu\partial v^A_\alpha} \d v^B_\nu\wedge\d
y^A\wedge\d^{m-1}x_\alpha -\frac{\partial^2L}{\partial y^B\partial
v^A_\alpha}\d y^B\wedge \d y^A\wedge\d^{m-1}x_\alpha
\nonumber
  \\  & &
\displaystyle + \frac{\partial^2L}{\partial v^B_\nu\partial v^A_\alpha}v^A_\alpha
\d v^B_\nu\wedge\d^mx  +
\left(\frac{\partial^2L}{\partial y^B\partial v^A_\alpha}v^A_\alpha
  -\derpar{L}{y^B}+
\frac{\partial^2L}{\partial x^\alpha\partial v^B_\alpha}
\right)\d y^B\wedge\d^mx
 \label{+}
\eea
(where $\d^{m}x = \d x^1 \wedge \ldots \wedge \d x^m$ and
\dst\d^{m-1}x^\alpha\equiv\inn\left(\derpar{}{x^\alpha}\right)\d^mx\)).
Locally, the regularity condition is equivalent to
\dst det\left(\frac{\partial^2 L} {\partial v^A_\alpha\partial
v^B_\nu}(\bar y)\right) \not= 0\),
for every $\bar y\in J^1\pi$.

The {\sl Lagrangian problem} associated with a Lagrangian system
 $\ls$ consists in finding sections $\phi\in\Gamma(M,E)$
(where $\Gamma(M,E)$ denotes the set of sections of $\pi$),
such that
 $$
 (j^1\phi)^*\inn (X)\Omega_\Lag=0 \quad ,\quad
 \mbox{\rm for every $X\in\vf (J^1\pi)$}
 $$
 In natural coordinates this is equivalent to demanding that
 $\phi$ satisfies the {\sl Euler-Lagrange equations}.
 The problem of finding these sections
 can be formulated equivalently as follows: to
 find the integral sections of a class of
 {\sl holonomic} $m$-vector fields $\{ {\cal X}_{\Lag}\}\subset\vf^m(J^1\pi)$,
 such that
 $$
 \inn ({\cal X}_{\Lag})\Omega_{\Lag}=0
 \quad , \quad
 \mbox{\rm for every ${\cal X}_\Lag\in\{ {\cal X}_{\Lag}\}$}
 $$
(Holonomic means that ${\cal X}_\Lag$
is integrable and its integral sections are holonomic. This is equivalent
to demanding that ${\cal X}_\Lag$ is integrable and {\sl semi-holonomic}, 
that is, it satisfies the condition $\inn({\cal X}_\Lag){\cal V} = 0$.
 {\sl Semi-holonomic} (not necessarily integrable)
 locally decomposable $m$-vector fields which are solution to these equations
 are called {\sl Euler-Lagrange $m$-vector fields} for $\ls$.

For the Hamiltonian formalism of field theories,
we take as the {\sl multimomentum bundle} the manifold
 $J^1\pi^*\equiv\Lambda_2^m\Tan^*E/\pi^*\Lambda^m\Tan^*M$,
 where $\Lambda_2^m\Tan^*E\equiv {\cal M}\pi$
 is the bundle of $m$-forms on
 $E$ vanishing by the action of two $\pi$-vertical vector fields.
 It is a bundle
 $\bar\tau^1=\pi\circ\tau^1\colon J^1\pi^*\to M$,
where $\tau^1\colon J^1\pi^*\to E$ is the natural projection.
Natural charts of coordinates in ${\cal M}\pi$ and $J^1\pi$
(adapted to the bundle structure, and such that
$\omega^*\equiv\bar\tau^{1*}\eta=\d x^1\wedge\ldots\wedge\d x^m\equiv\d x^m$)
are denoted by $(x^\alpha,y^A,p_A^\alpha,p)$ and $(x^\alpha,y^A,p_A^\alpha)$,
respectively.

As ${\cal M}\pi$ is a subbundle of $\Lambda^m\Tan^*E$
(the multicotangent bundle of $E$ of order $m$),
then ${\cal M}\pi$ is endowed with canonical forms:
the ``tautological form''
$\Theta\in\df^m({\cal M}\pi)$, and the multisymplectic form
 $\Omega:=-\d\Theta\in\df^{m+1}({\cal M}\pi)$.
They are known as the {\sl multimomentum Liouville $m$ and $(m+1)$-forms}.
 Their local expressions are
  $$
  \Theta = p_A^\alpha\d y^A\wedge\d^{m-1}x_\alpha+p\d^mx
  \quad , \quad
  \Omega = -\d p_A^\alpha\wedge\d y^A\wedge\d^{m-1}x_\alpha-\d p\wedge\d^mx
  $$

Now, if $\ls$ is a Lagrangian system,
 the {\sl extended Legendre map} associated with $\Lag$,
 $\widetilde{{\cal F}\Lag}\colon J^1\pi\to {\cal M}\pi$,
 is defined by:
  $ (\widetilde{{\cal F}\Lag}\bar y))(\moment{Z}{1}{m}):=
 (\Theta_{\Lag})_{\bar y}(\moment{\bar Z}{1}{m})$,
 for $\bar y\in J^1\pi$, where $\moment{Z}{1}{m}\in\Tan_{\pi^1(\bar y)}E$, and
 $\moment{\bar Z}{1}{m}\in\Tan_{\bar y}J^1\pi$ are such that
 $\Tan_{\bar y}\pi^1\bar Z_\alpha=Z_\alpha$.
 Then, using the natural projection
 $\mu \colon {\cal M}\pi\to J^1\pi^*$,
 we define the {\sl restricted Legendre map} associated with $\Lag$ as
 ${\cal F}\Lag :=\mu\circ\widetilde{{\cal F}\Lag}$.
Their local expressions are
 $$
 \begin{array}{ccccccc}
 \widetilde{{\cal F}\Lag}^*x^\nu = x^\nu &\quad\ , \ \quad&
 \widetilde{{\cal F}\Lag}^*y^A = y^A &\quad\  , \quad&
 \widetilde{{\cal F}\Lag}^*p_A^\nu =\derpar{L}{v^A_\nu}
 &\quad\ , \quad&
 \widetilde{{\cal F}\Lag}^*p =L-v^A_\nu\derpar{L}{v^A_\nu}
 \\
 {\cal F}\Lag^*x^\nu = x^\nu &\quad\ , \ \quad&
 {\cal F}\Lag^*y^A = y^A &\quad\  , \quad&
 {\cal F}\Lag^*p_A^\nu =\derpar{L}{v^A_\nu} & &
 \end{array}
 $$
 We have that
 $\widetilde{{\cal F}\Lag}^*\Theta=\Theta_{\Lag}$,
 and $\widetilde{{\cal F}\Lag}^*\Omega=\Omega_{\Lag}$.

 $\ls$ is a {\sl regular}
 Lagrangian system if ${\cal F}\Lag$ is a local diffeomorphism
 (this definition is equivalent to that given above).
 Elsewhere $\ls$ is a {\sl singular} Lagrangian system.
 As a particular case, $\ls$ is a {\sl hyper-regular}
 Lagrangian system if ${\cal F}\Lag$ is a global diffeomorphism.
 A singular Lagrangian system $\ls$ is {\sl almost-regular} if
 ${\cal P}:={\cal F}\Lag (J^1\pi)$ is a closed submanifold of $J^1\pi^*$,
 ${\cal F}\Lag$ is a submersion onto its image, and
 for every $\bar y\in J^1\pi$, the fibres
 ${\cal F}\Lag^{-1}({\cal F}\Lag (\bar y))$
 are connected submanifolds of $J^1\pi$.

 If $\ls$ is an almost-regular Lagrangian system then
 ${\cal P}$ is a fibre bundle over $E$ and $M$ (the natural
projections are denoted by $\tau_0^1\colon{\cal P}\to E$ and
$\bar\tau_0^1:=\pi\circ\tau_0^1\colon{\cal P}\to M$) and the
$\mu$-transverse submanifold
 $\tilde{\cal P} = \widetilde{\cal F \Lag}(J^1\pi) \hookrightarrow{\cal M}\pi$ 
is diffeomorphic to ${\cal P}$ (and we denote by
  $\tilde\jmath_0\colon\tilde{\cal P}\hookrightarrow{\cal M}\pi$
the natural imbedding).
 This diffeomorphism is denoted
 $\tilde\mu\colon\tilde{\cal P}\to{\cal P}$,
 and it is just the restriction of the projection $\mu$ to $\tilde{\cal P}$.
 Then, taking $\tilde h:=\tilde\mu^{-1}$,
 we define the Hamilton-Cartan $(m+1)$-form
  $\Omega^0_h=(\tilde\jmath_0\circ\tilde h)^*\Omega$,
 which verifies that
 ${\cal F}\Lag_0^*\Omega^0_h=\Omega_{\Lag}$
(where ${\cal F}\Lag_0$ is the restriction map of
${\cal F}\Lag$ onto ${\cal P}$).
 Then $\tilde h$ is called a {\sl Hamiltonian section}, and
 $\hso$ is the {\sl Hamiltonian system}
 associated with the almost-regular Lagrangian system $\ls$ (see \cite{LMM-96}).

If $\ls$ is a hyper-regular Lagrangian system, then ${\cal
P}=J^1\pi^*$, and the construction is the same. In addition,
 $\widetilde{{\cal F}\Lag}(J^1\pi)$ is a
 1-codimensional embedded submanifold of ${\cal M}\pi$,
 which is transverse to the projection $\mu$, and is diffeomorphic to
 $J^1\pi^*$. This diffeomorphism is $\mu^{-1}$, when $\mu$ is
 restricted to $\widetilde{{\cal F}\Lag}(J^1\pi)$, and coincides with the map
 $h:=\widetilde{{\cal F}\Lag}\circ{\cal F}\Lag^{-1}$,
 when it is restricted onto its image.
 $h$ is the {\sl Hamiltonian section} in this case, and
the associated Hamiltonian system is denoted by $(J^1\pi^*,\Omega_h)$,
where $\Omega_h=h^*\Omega$.
 In a local chart of natural coordinates,
 the Hamiltonian section is specified by a {\sl local Hamiltonian function}
 $H\in\Cinfty (U)$, $U\subset J^1\pi^*$, such that
 $h(x^\alpha,y^A,p^\alpha_A)\equiv
 (x^\alpha,y^A,p^\alpha_A,p=-H)$, where
  $$
 H(x^\alpha,y^A,p^\alpha_A)=
 ({\cal F}\Lag^{-1})^*\left(v^A_\alpha\derpar{L}{v^A_\alpha}-L\right)=
 p^\alpha_A({\cal F}\Lag^{-1})^*v_\alpha^A- ({\cal F}\Lag^{-1})^*L
 $$
and $\Omega_h=-\d p_A^\alpha\wedge\d y^A\wedge\d^{m-1}x_\alpha+\d H\wedge\d^mx$.

 The {\sl Hamiltonian problem} associated with the Hamiltonian
 system $\hso$ (for $(J^1\pi^*,\Omega_h)$ is analogous),
 consists in finding sections $\psi_o\in\Gamma(M,{\cal P})$
such that
 $$
 \psi_o^*\inn (X_0)\Omega^0_h=0 \quad , \quad
 \mbox{\rm  for every $X_0\in\vf ({\cal P})$}
 $$
 As in the Lagrangian case, these sections are the integral sections of a class of
 integrable and $\bar\tau^1_0$-transverse $m$-vector fields
 $\{ {\cal X}_{{\cal H}_o}\}\subset\vf^m({\cal P})$ satisfying that
 $$
 \inn ({\cal X}_{{\cal H}_o})\Omega^0_h=0 \quad ,   \quad
 \mbox{\rm for every ${\cal X}_{{\cal H}_o}\in\{ {\cal X}_{{\cal H}_o}\}.$}
 $$
$m$-vector fields satisfying these conditions
(but not necessarily integrable) are called
{\sl Hamilton-De Donder-Weyl $m$-vector fields} for $\hso$.

\subsection{Lagrangian and Hamiltonian algorithms}

Let $\ls$ be a Lagrangian system. If $\nabla$ is an Ehresmann
connection in the fibration $\bar{\pi}^{1} \colon J^1\pi \to M$,
let ${\cal Y}^{\nabla}_{\eta}$ be the corresponding $m$-vector
field on $J^1\pi$. Then, we have:

\begin{prop}
\label{pb1}
The Poincar\'e-Cartan $(m+1)$-form may be written as
$$
\Omega_{\Lag}= \omega \wedge (\gamma_{\Lag})^{\nabla}_{\eta} +
\Omega_{\Lag}^{\nabla},
$$
where $(\gamma_{\Lag})_{\eta}^{\nabla}=\inn({\cal Y}^{\nabla}_{\eta})\Omega_{\Lag}
\in\df^1(J^1\pi)$,
and $\Omega_{\Lag}^{\nabla}$ is a $(m+1)$-form on $J^1\pi$ of
bidegree $(m-1,2)$ with respect to the connection $\nabla$.
\end{prop}
\proof If $y \in J^1\pi$ and $v_{1}, v_{2}, v_{3} \in
V_{y}(\bar{\pi}^{1})$ ($V(\bar{\pi}^1)$ being the vertical bundle
of $\bar{\pi}^1$) then, from (\ref{+}) we have that $i(v_{1}
\wedge v_{2} \wedge v_{3}) \Omega_{\Lag}(y) = 0$. Thus,
the result follows from Propositions \ref{gamma} and \ref{gamma1}.
\qed

If $\ls$ is a hyperregular Lagrangian system (the regular case is analogous)
${\cal F}\Lag$ is a global diffeomorphism.
Moreover, if $\Omega_{h}$ is the Hamilton-Cartan
$(m+1)$-form on $J^1\pi^*$, then
\beq
\label{eb1}
{\cal F}\Lag^{*}\Omega_{h} = \Omega_{\Lag}.
\eeq
Furthermore, as ${\cal F}\Lag$ is a global diffeomorphism,
the connection $\nabla$ induces a connection $\nabla^{*}$ in the
fibration $\bar{\tau}^{1} \colon J^1\pi^* \to M$ in such a way that
\beq
\label{eb2}
{\cal F}\Lag_{*}{\cal Y}^{\nabla}_{\eta} = {\cal Y}_{\eta}^{\nabla^*},
\eeq
where ${\cal Y}^{\nabla^*}_{\eta}$ is the $m$-vector field on
$J^1\pi^*$ associated with $\nabla^{*}$ and the volume form $\eta$.
 Thus, from (\ref{eb1}), (\ref{eb2}) and Proposition \ref{pb1}, we
 obtain:

\begin{prop}
\label{pb1bis}
The Hamilton-Cartan $(m+1)$-form may be written as
$$
\Omega_{h}=\Omega_{h}^{\nabla^*}+\omega^{*}\wedge(\gamma_{h})^{\nabla^{*}}_{\eta},
$$
where $(\gamma_{h})^{\nabla^*}_{\eta}=\inn({\cal
Y}^{\nabla^*}_{\eta})\Omega_{h}$, and $\Omega_{h}^{\nabla^*}$ is a
$(m+1)$-form on $J^1\pi^*$ of bidegree $(m-1,2)$ with respect to
the connection $\nabla^*$.
\end{prop}

Furthermore, we may prove the following result:

\begin{prop}
\label{pb2}
If $\ls$ is a regular Lagrangian system, then the $(m+1)$-forms
$\Omega_{\Lag}^{\nabla}$ and $\Omega_{h}^{\nabla^*}$ are $1$-nondegenerate.
\end{prop}
\proof As ${\cal F}\Lag$ is a diffeomorphism and ${\cal
F}\Lag^{*}\Omega_{h}^{\nabla^*} = \Omega_{\Lag}^{\nabla}$, it
suffices to prove that $\Omega_{h}^{\nabla^*}$ is
$1$-nondegenerate. The local expression of $\Omega_{h}^{\nabla^*}$
is
\[
\Omega_{h}^{\nabla^*} = -dp_{A}^{\alpha} \wedge dy^{A} \wedge
d^{m-1}x_{\alpha} + \theta \wedge d^{m}x,
\]
$\theta$ being a $1$-form such that
$\displaystyle\theta\left(\frac{\partial}{\partial x_{\alpha}}\right)=0$,
for every $\alpha$. As a consequence,
\beq
\label{eb3}
\displaystyle \inn (\frac{\partial}{\partial x_{\beta}})
\Omega_{h}^{\nabla^*}= \displaystyle -\sum_{A, \alpha ; \alpha \neq \beta}
dp_{A}^{\alpha} \wedge dy^{A} \wedge d^{m-2}x_{\alpha \beta}
-\theta\left(\frac{\partial}{\partial y^{A}}\right)dy^{A} \wedge
d^{m-1}x_{\beta} \\[7pt] - \displaystyle
\theta\left(\frac{\partial}{\partial p_{A}^{\alpha}}\right)
dp_{A}^{\alpha} \wedge d^{m-1}x_{\beta} \eeq \beq \label{eb4} \inn
\left(\frac{\partial}{\partial y^{A}}\right) \Omega_{h}^{\nabla^*}
= \sum_{\alpha} dp_{A}^{\alpha} \wedge d^{m-1}x_{\alpha} +
\theta\left(\frac{\partial}{\partial y^{A}}\right) d^m x,
\eeq
\beq
\label{eb5} \inn \left(\frac{\partial}{\partial
p_{A}^{\alpha}}\right) \Omega_{h}^{\nabla^*} = - dy^{A} \wedge
d^{m-1}x_{\alpha} + \theta\left(\frac{\partial}{\partial
p_{A}^{\alpha}}\right) d^m x.
\eeq
Thus, if $X = \displaystyle
\lambda_{\beta} \frac{\partial}{\partial x_{\beta}} + \mu^{A}
\frac{\partial}{\partial y^{A}} + \nu^{\alpha}_{A}
\frac{\partial}{\partial p_{A}^{\alpha}}$ is a local vector field
such that $\inn (X)\Omega_{h}^{\nabla^*} = 0$ then, from
(\ref{eb3}), it follows that $\lambda_{\beta} = 0$, for every
$\beta$, which implies that (see (\ref{eb4}) and (\ref{eb5}))
\[
\mu^{A} dp_{A}^{\alpha} \wedge d^{m-1}x_{\alpha} + \mu^{A}
\theta(\frac{\partial}{\partial y^{A}}) d^{m}x
- \nu^{\alpha}_{A} dy^{A} \wedge d^{m-1}x_{\alpha} +
\nu^{A}_{\alpha} \theta(\frac{\partial}{\partial p_{A}^{\alpha}}) d^{m}x = 0.
\]
Therefore, $\mu^{A} = 0$ and $\nu^{\alpha}_{A} = 0$, for every $A$ and
$\alpha$, that is, $X = 0$.
\qed

If the Lagrangian is regular, then from Propositions
\ref{pr1} and \ref{pb2}, we obtain that
$(T_{y}^{\perp}J^1\pi)^{\nabla}_{\Omega_{\Lag}} = \{ 0\}$, for
every $y \in J^1\pi$. Thus, there exist locally decomposable
$m$-vector fields ${\cal X}_{\Lag}$ on $J^1\pi$ such that
\[
i({\cal X}_{\Lag})\omega = 1, \makebox[.4cm]{} i({\cal
X}_{\Lag})\Omega_{\Lag} = 0.
\]
Moreover, we have
\begin{prop}
\label{p10} If $(J^1\pi, \Omega_{\Lag})$ is a regular Lagrangian
system and ${\cal X}_{\Lag}$ is a locally decomposable $m$-vector
field on $J^1\pi$ such that $i({\cal X}_{\Lag})\omega = 1$ and
$i({\cal X}_{\Lag})\Omega_{\Lag} = 0$ then ${\cal X}_{\Lag}$ is an
Euler-Lagrange $m$-vector field for $\Lag$.
\end{prop}
\proof We must prove that ${\cal X}_{\Lag}$ is semi-holonomic,
that is, $i({\cal X}_{\Lag}){\cal V} = 0.$ For this purpose, we
consider local fibred coordinates $(x^{\alpha}, y^{A},
v^{A}_{\alpha})$ on $J^1\pi$. Then, since $i({\cal
X}_{\Lag})\omega = 1$, it follows that
\[
{\cal X}_{\Lag} = \Lambda_{\alpha=1}^{m}\left(\displaystyle
\frac{\partial}{\partial x^{\alpha}} + \Gamma_{\alpha}^{A}
\frac{\partial}{\partial y^{A}} + \Gamma_{\alpha\beta}^{A}
\frac{\partial}{\partial v_{\beta}^{A}}\right)
\]
with $\Gamma_{\alpha}^{A}$ and $\Gamma_{\alpha\beta}^{A}$ local
real functions on $J^1\pi$. Furthermore, from (\ref{+}), we deduce
that
\[
0 = (i({\cal X}_{\Lag})\Omega_{\Lag})(\displaystyle
\frac{\partial}{\partial v_{\nu}^{B}}) = (-1)^m i({\cal
X}_{\Lag})(i(\displaystyle \frac{\partial}{\partial
v_{\nu}^{B}})\Omega_{\Lag})
\]
\[
= (-1)^{m+1}(\Gamma_{\alpha}^A - v_{\alpha}^A) \displaystyle
\frac{\partial^2 L}{\partial v_{\alpha}^A \partial v_{\nu}^B}, \;
\; \mbox{ for all } B \mbox{ and } \nu .
\]
Therefore, using the fact that $\Lag$ is regular, we conclude that
\[
\Gamma_{\alpha}^A = v_{\alpha}^A, \; \; \mbox{ for all } A \mbox{
and } \alpha,
\]
which implies that ${\cal X}_{\Lag}$ is semi-holonomic.
\qed

Hence, if $\ls$ is a regular Lagrangian system, then
 the existence of classes of Euler-Lagrange $m$-vector fields for $\Lag$
 is assured in $J^1\pi$. In the same way,
for the Hamiltonian formalism, the existence of
 Hamilton-De Donder-Weyl $m$-vector fields is assured
 everywhere in $J^1\pi^*$ (note that if ${\cal X}_{\Lag}$ is
 an Euler-Lagrange $m$-vector field for $\Lag$ then
 $({\cal F}{\cal L})_{*} {\cal X}_{\Lag}$ is a Hamilton-De Donder-Weyl
 $m$-vector field on $J^1\pi^*$).
 In both cases, the solution is not unique.

For singular (almost-regular) Lagrangian systems,
the existence of Euler-Lagrange $m$-vector fields
is not assured except perhaps
on some submanifold $S_f\hookrightarrow J^1\pi$,
where the solution is not unique.
In order to find this submanifold
we apply the algorithm developed in
Section \ref{ca} to the system $\ls$,
by doing the identifications
$\kappa\colon F\to M$ with $\bar\pi^1\colon J^1\pi\to M$,
and $\Omega$ with $\Omega_\Lag$.
Thus we obtain obtain a sequence
\beq
 \cdots \stackrel{j^i_{i+1}}{\hookrightarrow} N_i
 \stackrel{j^{i-1}_i}{\hookrightarrow} \cdots
 \stackrel{j^1_2}{\hookrightarrow} N_1
 \stackrel{j_1}{\hookrightarrow} N_0\equiv J^1\pi\; .
 \label{seqsubmanlag}
 \eeq
which, in the best of cases stabilizes in the final constraint
submanifold $N_f$ where there exist $m$-vector fields ${\cal
X}^{N_{f}}$ on $N_{f}$, solution to the equations
\beq \label{24'}
(\inn ({\cal X}^{N_{f}})\Omega_{\Lag})\vert_{N_{f}} = 0,
\makebox[.4cm]{} (\inn ({\cal X}^{N_{f}})\omega)\vert_{N_{f}} = 1.
\eeq
But ${\cal X}^{N_{f}}$ will not be, in general, an
Euler-Lagrange $m$-vector field on $N_{f}$ (that is, it is not
semi-holonomic), and, in addition, ${\cal X}^{N_{f}}$ will not in
general be an integrable $m$-vector field. The problem of finding
integrable Euler-Lagrange $m$-vector fields (i.e., holonomic) is
discussed and solved in the next Section.

Now, we consider the Hamiltonian system $\hso$. Let $\nabla_0^*$
be a connection in the bundle $\bar\tau^1_0\colon{\cal P}\to M$
and denote by ${\cal Y}^{\nabla_0^*}_{\eta}$ the corresponding
$m$-vector field on ${\cal P}$ associated with $\nabla_0^{*}$ and
$\eta$. Then, we have:

\begin{prop}
\label{pb1bbis}
The Hamilton-Cartan $(m+1)$-form may be written as
$$
\Omega_{h}^0=
\Omega_{h}^{\nabla_0^*}+\omega_0^{*}\wedge(\gamma_{h})^{\nabla_0^*}_{\eta},
$$
where $\omega_0^{*}=\bar\tau^{1*}_0\eta$,
$(\gamma_{h})^{\nabla_0^*}_{\eta}=\inn({\cal
Y}^{\nabla_0^*}_{\eta})\Omega_{h}$, and $\Omega_{h}^{\nabla_ 0^*}$
is a $(m+1)$-form on ${\cal P}$ of bidegree $(m-1,2)$ with respect
to the connection $\nabla_0^*$.
\end{prop}
\proof If $\bar{y} = {\cal F}{\cal L}_{0}(y) \in {\cal P}$, with
$y \in J^1\pi$, and $\bar{v}_{1}, \bar{v}_{2}, \bar{v}_{3} \in
V_{y}(\bar{\tau}_{0}^1)$ then, since ${\cal F}{\cal L}_{0}: J^1\pi
\to {\cal P}$ is a submersion and $\bar{\tau}_{0}^1 \circ {\cal
F}{\cal L}_{0} = \bar{\pi}^{1}$, it follows that there exist
$v_{1}, v_{2}, v_{3} \in V_{y}(\bar{\pi}^1)$ such that
\[
(T_{y}{\cal F}{\cal L}_{0})(v_{i}) = \bar{v}_{i}, \; \; \mbox{ for
} i \in \{1, 2, 3\}.
\]
Thus, using that $({\cal F}{\cal L}_{0})^* \Omega_{h}^{0} =
\Omega_{\Lag}$, we deduce that
\[
i(\bar{v}_{1}\wedge\bar{v}_{2}\wedge\bar{v}_{3})\Omega_{h}^{0}(\bar{y})
= 0.
\]
This proves the result.
\qed

Hamilton-De Donder-Weyl $m$-vector fields do not exist, in
general,
 in ${\cal P}$, and then we must apply the algorithmic procedure
developed in Section \ref{ca} to the system $\hso$,
by doing the identifications
$\kappa\colon F\to M$ with $\bar\tau^1\vert_{\cal P}\colon{\cal P}\to M$,
and $\Omega$ with $\Omega_h^0$.
Thus we obtain a sequence
\beq
 \cdots \stackrel{j^i_{i+1}}{\hookrightarrow} P_i
 \stackrel{j^{i-1}_i}{\hookrightarrow} \cdots
 \stackrel{j^1_2}{\hookrightarrow} P_1
 \stackrel{j_1}{\hookrightarrow} P_0\equiv {\cal P}\; .
 \label{seqsubmanham}
 \eeq
which, in the best of cases stabilizes in the final constraint
submanifold $P_f$  of ${\cal P}$ where there exist $m$-vector
fields ${\cal X}^{P_{f}}$ on $P_{f}$, solution to the equations
\beq
\label{25'} (\inn ({\cal X}^{P_{f}})\Omega_h^0)\vert_{P_{f}}
= 0, \makebox[.4cm]{} (\inn ({\cal
X}^{P_{f}})\omega_{0}^*)\vert_{P_{f}} = 1.
\eeq
 Of course the solution ${\cal X}^{P_{f}}$ is not unique.

\begin{remark}\label{equiv}{\rm
The Lagrangian and Hamiltonian pre-multisymplectic algorithms are
equivalent in the following sense: at every level $j$ of the
Lagrangian and Hamiltonian algorithms, the submanifolds of the
sequences (\ref{seqsubmanlag}) and (\ref{seqsubmanham}) are ${\cal
F}\Lag$-related, that is, ${\cal F}\Lag(N_j)=P_j$ and ${\cal
F}{\cal L}_{j} = {\cal F}{\cal L}_{|N_{j}}: N_{j} \to P_{j}$ is a
submersion such that ${\cal F}{\cal L}_{j}^{-1}({\cal F}{\cal
L}_{j}(x_{j})) = {\cal F}{\cal L}_{0}^{-1}({\cal F}{\cal
L}_{0}(x_{j}))$, for $x_{j} \in N_{j}$. Moreover, if $N_{f}$ is
the final constraint submanifold (in the Lagrangian level) and
${\cal X}^{N_{f}}$ is a locally decomposable $m$-vector field on
$N_{f}$ such that equations (\ref{24'}) hold and, in addition,
${\cal X}^{N_{f}}$ is ${\cal F}{\cal L}_{f}$-projectable to an
$m$-vector field ${\cal X}^{P_{f}}$ on $P_f$ then ${\cal
X}^{P_{f}}$ is locally decomposable and equations (\ref{25'})
hold. Conversely, if ${\cal X}^{P_{f}}$ is a locally decomposable
$m$-vector field on $P_{f}$ satisfying equations (\ref{25'}) and
${\cal X}^{N_{f}}$ is a locally decomposable $m$-vector field on
$N_{f}$ which is ${\cal F}{\cal L}_{f}$-projectable on ${\cal
X}^{P_{f}}$ then ${\cal X}^{N_f}$ satisfies equations (\ref{24'})
(see \cite{LMM-96,LMMa-2002} for a detailed discussion on this
topic).}
\end{remark}

Finally, the Hamilton-De Donder-Weyl $m$-vector fields
${\cal X}^{P_f}$ are not integrable, in general. In
fact, if we have that
${\cal X}^{P_f} = X^{P_f}_{1}
\wedge \ldots \wedge  X^{P_f}_{m}$,
where $X^{P_f}_{\alpha}$ are (local) vector
fields on $P_{f}$, for all $\alpha$, and
$\{ X^{P_f}_{1},\ldots ,X^{P_f}_{m},\bar{Z}_{1}, \ldots, \bar{Z}_{p} \}$
is a local basis of the vector bundle $TP_{f} \to P_{f}$ then
\[
[X^{P_f}_{\alpha},X^{P_f}_{\beta}] =
\bar{f}_{\alpha \beta}^{\gamma}X^{P_f}_{\gamma}+\bar{\zeta}_{\alpha \beta}^{l}
\bar{Z}_{l}
\]
for some functions $\bar{f}^{\gamma}_{\alpha \beta}$ and
$\bar{\zeta}^{l}_{\alpha \beta}$ on $P_{f}$.
Therefore, we must apply the integrability algorithm of Section \ref{intalg},
and we obtain a sequence
$\ldots \subseteq {\cal J}_{i} \subseteq \ldots \subseteq {\cal
J}_{1} \subseteq P_{f}$,
such that ${\cal J}_{i}$ is a non-empty (closed) submanifold of
$S_{f}$, with
\beann
{\cal J}_{1} &=& \{ y \in P_{f} \ \vert\ \bar{\zeta}_{\alpha \beta}^{l}(y)
= 0 \}
\\
{\cal J}_{i} &=& \{ y \in {\cal J}_{i-1} \ \vert\ {\cal
X}^{P_f}(y) \in \Lambda^{m}\Tan_{y}{\cal J}_{i-1} \}, \; \; \;
\mbox{for} \; \; i \geq 2. \eeann

In the best cases, there exists an integer $i$ such that ${\cal
J}_{i+1} = {\cal J}_{i}$. Then, ${\cal J}_{f} = {\cal J}_{i+1} =
{\cal J}_{i}$ is a submanifold of $P_{f}$, and ${\cal X}^{{\cal
J}_{f}} = ({\cal X}^{P_{f}})\vert_{{\cal J}_{f}}$ is an integrable
Hamilton-De Donder-Weyl $m$-vector field in ${\cal J}_{f}$.

\subsection{Almost-regular Lagrangians and integrable
Euler-Lagrange $m$-vector fields}

Let $(J^1\pi, \Omega_{\Lag})$ be an almost-regular Lagrangian
system, and $N_f$ the final constraint submanifold (in the
Lagrangian setting). Then, there exists a locally decomposable
$m$-vector field ${\cal X}^{N_{f}}$ on $N_{f}$ such that
\[
(\inn ({\cal X}^{N_{f}})\Omega_{\Lag})\vert_{N_{f}} = 0,
\makebox[.4cm]{} (\inn ({\cal X}^{N_{f}})\omega)\vert_{N_{f}} = 1.
\]
But, in general, ${\cal X}^{N_{f}}$ is not an Euler-Lagrange
$m$-vector field on $N_{f}$ and, in addition, ${\cal X}^{N_{f}}$
will not in general be an integrable $m$-vector field.

In order to solve these problems, first we construct a submanifold
$S_{f}$ of $N_{f}$ where there exists a locally decomposable
$m$-vector field ${\cal X}^{S_{f}}$ such that
\[
(\inn ({\cal X}^{S_{f}})\Omega_{\Lag})\vert_{S_{f}} = 0,
\makebox[.3cm]{} (\inn ({\cal X}^{S_{f}})\omega)\vert_{S_{f}} = 1,
\makebox[.3cm]{} (\inn ({\cal X}^{S_{f}}){\cal V})\vert_{S_{f}} = 0.
\]
In fact, from the above discussion we know that we can choose the
$m$-vector field ${\cal X}^{N_{f}}$ on $N_{f}$ such that it
projects via ${\cal F}\Lag_{f}$ (the restriction of ${\cal F}\Lag$
to $N_f$) onto an $m$-vector field ${\cal X}^{P_{f}}$ on $P_{f}$.
Then, we consider the subset $S_{f}$ of $N_{f}$ defined by \beq
\label{esef}
 S_{f} = \{ x \in N_{f} / (\inn ({\cal X}^{N_{f}}){\cal V})(x) =
 0\}.
\eeq In \cite{LMMa-2002} (see also \cite{LMM-96}), it was proved
that \beq \label{tanlev} (\inn({\cal X}^{N_{f}}){\cal V})(x) \in
Ker \Tan_{x}({\cal F}\Lag_{f}) = Ker \Tan_{x}({\cal F}\Lag_{0})
\eeq and that for every $x \in N_{f}$, $S_{f} \cap {\cal
F}\Lag_{f}^{-1}({\cal F}\Lag_{f}(x)) = S_{f} \cap {\cal
F}\Lag_{0}^{-1}({\cal F}\Lag_{0}(x))$ is a single point in
$S_{f}$.

The above result allows us to introduce a well-defined map $s_{f}:
P_{f} \to N_{f}$ such that
\[
S_{f} = s_ {f}(P_{f}), \makebox[.4cm]{} {\cal F}\Lag_{f} \circ
s_{f} = Id.
\]
Thus, $s_{f}: P_{f} \to N_{f}$ is a global section of the
submersion ${\cal F}\Lag_{f}: N_{f} \to P_{f}$ and, therefore,
$S_{f}$ is an embedded submanifold of $N_{f}$ and the map $s_{f}:
P_{f} \to S_{f}$ is a diffeomorphism (for more details, see
\cite{LMM-96,LMMa-2002}).

Now, defining the $m$-vector field ${\cal X}^{S_{f}}$ on
$S_{f}$ by
${\cal X}^{S_{f}} = (\Lambda^{m}\Tan s_{f}) \circ {\cal X}^{P_{f}}$,
then we have \cite{LMMa-2002}:

\begin{teor}
 ${\cal X}^{S_{f}}$ is an Euler-Lagrange
$m$-vector field on $S_{f}$ for the Lagrangian $\Lag$, that is,
${\cal X}^{S_{f}}$ is a locally decomposable $m$-vector field on
$S_{f}$ and
\[
(\inn ({\cal X}^{S_{f}})\Omega_{\Lag})\vert_{S_{f}} = 0,
\makebox[.3cm]{} (\inn ({\cal X}^{S_{f}})\omega)\vert_{S_{f}} = 1,
\makebox[.3cm]{} (\inn ({\cal X}^{S_{f}}){\cal V})\vert_{S_{f}} = 0.
\]
\label{lmma}
\end{teor}

\vspace{-1cm}

Next, we give a local description of the submanifold $S_{f}$ and
of the Euler-Lagrange $m$-vector field ${\cal X}^{S_{f}}$ on
$S_{f}$. Since $\Lag$ is almost-regular, it follows that the rank
of the partial Hessian matrix $\left(\displaystyle
\frac{\partial^2L}{\partial v^A_{\alpha}
\partial v^B_{\beta}}\right)$
 is constant. Let
$rank \left(\displaystyle \frac{\partial^2 L}{\partial v^A_{\alpha}
\partial v^B_{\beta}}\right)=pm + q$, with $0 \leq p \leq n-1$ and
$0 \leq q \leq m$, and assume that the first $pm + q$ rows of this
matrix are independent. Denote by $\widetilde{V(\pi^1)} \to
J^1\pi$ the vector subbundle of the vertical bundle ${V(\pi^{1})}
\to J^1\pi$ of $\pi^1: J^1\pi \to M$ generated by the local vector
fields
\[
\left\{\displaystyle \frac{\partial}{\partial v^A_{\alpha}},
\frac{\partial}{\partial v^{p+1}_{1}}, \ldots,
\frac{\partial}{\partial v^{p+1}_{q}}\right\}, \; \; \; \mbox{for} \; 1
\leq A \leq p \;\; \mbox{and} \;\; 1 \leq \alpha \leq m.
\]
Then, there exist sections $\{ X^{p+1}_{q+1}, \ldots, X^{p+1}_{m},
X^{A}_{\alpha}\}$, with $p+2 \leq A \leq n$ and $1 \leq \alpha
\leq m$, of the vector bundle $\widetilde{V(\pi^{1})} \to J^1\pi$
such that $\{ W^{p+1}_{q+1}, \ldots, W^{p+1}_{m},
W^{A}_{\alpha}\}$, with $p+2 \leq A \leq n$ and $1 \leq \alpha
\leq m$, is a local basis of $Ker (\Tan ({\cal F}\Lag_{0}))$,
where \beq \label{w} W^{p+1}_{\beta} = \displaystyle
\frac{\partial}{\partial v^{p+1}_{\beta}} + X^{p+1}_{\beta},
\makebox[.4cm]{} W^{A}_{\alpha} = \displaystyle
\frac{\partial}{\partial v^{A}_{\alpha}} + X^{A}_{\alpha}, \quad
(p+2\leq A\leq n,\ 1\leq\alpha\leq m,\ q+1\leq\beta\leq m).
 \eeq
Now, suppose that
${\cal X}^{N_{f}} = {\cal X}_{1}^{N_{f}} \wedge \ldots \wedge {\cal
X}_{m}^{N_{f}}$, with
\[
{\cal X}_{\alpha}^{N_{f}} = \left(\displaystyle
\frac{\partial}{\partial x^{\alpha}} + \Gamma_{\alpha}^{A}
\frac{\partial}{\partial y^{A}} + \Gamma_{\alpha \beta}^{A}
\frac{\partial}{\partial v^{A}_{\beta}}\right)\Big\vert_{N_{f}}, \; \;
\mbox{for} \; \; \alpha \in \{1, \ldots, m\}.
\]
Then, using that ${\cal X}^{N_{f}}$ is ${\cal
F}\Lag_{f}$-projectable, it follows that the functions
$\Gamma^{A}_{\alpha}$ are constant on the fibers of ${\cal
F}\Lag_{f}: N_{f} \to P_{f}$. But, as ${\cal F}\Lag_{f}^{-1}({\cal
F}\Lag_{f}(x)) = {\cal F}\Lag_{0}^{-1}({\cal F}\Lag_{0}(x))$, for
every $x \in N_{f}$ (see Remark \ref{equiv}), we obtain that
\beq
\label{w1}
\begin{array}{lcr}
W^{p+1}_{\gamma}(\Gamma_{\alpha}^{A}) = 0, && \gamma \in \{q+1,
\ldots, m\} \\
W^{p+1+i}_{\gamma}(\Gamma^{A}_{\alpha}) = 0, && i \in \{1, \ldots,
n-p-1 \}, \; \; \gamma \in \{1, \ldots, m\}.
\end{array}
\eeq
Furthermore
\[
\inn ({\cal X}^{N_{f}}){\cal V} = \displaystyle
(\Gamma^{A}_{\alpha} - v^{A}_{\alpha}) \frac{\partial}{\partial
v^{A}_{\alpha}}.
\]
Thus, from (\ref{tanlev}) and (\ref{w}), we have that \beq
\label{w2} \inn({\cal X}^{N_{f}}){\cal V} = (\Gamma^{p+1}_{\gamma}
- v^{p+1}_{\gamma}) W^{p+1}_{\gamma} + (\Gamma^{p+1+i}_{\gamma} -
v^{p+1+i}_{\gamma}) W^{p+1+i}_{\gamma}. \eeq Note that the
functions \beq \label{zeta}
\begin{array}{lcr}
\zeta^{p+1}_{\gamma} = \Gamma_{\gamma}^{p+1} - v^{p+1}_{\gamma},
&&
\gamma \in \{q+1, \ldots, m\} \\
\zeta^{p+1+i}_{\bar{\gamma}} = \Gamma^{p+1+i}_{\bar{\gamma}} -
v^{p+1+i}_{\bar{\gamma}}, && i \in \{1, \ldots, n-p-1 \}, \; \;
\bar{\gamma} \in \{1, \ldots, m\}.
\end{array}
\eeq are independent on $N_{f}$. In fact (see (\ref{w}),
(\ref{w1}) and (\ref{zeta}))
$$
\begin{array}{lcr}
W^{p+1}_{\gamma}(\zeta_{\gamma'}^{p+1}) = -\delta_{\gamma
\gamma'}, && W^{p+1}_{\gamma}(\zeta_{\bar{\gamma}}^{p+1+i}) = 0,  \\
W^{p+1+i}_{\bar{\gamma}}(\zeta^{p+1}_{\gamma}) = 0, &&
W^{p+1+i}_{\bar{\alpha}}(\zeta_{\bar{\gamma}}^{p+1+j}) =
-\delta_{ij}\delta_{\bar{\alpha} \bar{\gamma}}.
\end{array}
$$
 Moreover, using (\ref{esef}) and (\ref{w2}), we conclude that
$\{\zeta_{\gamma}^{p+1}, \zeta_{\bar{\gamma}}^{p+1+i}\}$, with
$\gamma \in \{1, \ldots, m\}$, $i \in \{1, \ldots, n-p-1\}$ and
$\bar{\gamma} \in \{1, \ldots, m\}$, is a set of local independent
constraint functions defining $S_{f}$ as a submanifold of $N_{f}$,
that is,
\[
S_{f} = \{ x \in N_{f} / (\Gamma_{\gamma}^{p+1} -
v^{p+1}_{\gamma})(x) = 0, \makebox[.3cm]{}
(\Gamma_{\bar{\gamma}}^{p+1+i} - v^{p+1+i}_{\bar{\gamma}})(x) =
0\}.
\]
Finally, a direct calculation proves that the Euler-Lagrange
$m$-vector field  ${\cal X}^{S_{f}}$ on $S_{f}$ is given by
${\cal X}^{S_{f}} = {\cal X}_{1}^{S_{f}} \wedge \ldots \wedge {\cal
X}_{m}^{S_{f}}$, with
\[
{\cal X}_{\alpha}^{S_{f}} = ({\cal X}_{\alpha}^{N_{f}} + {\cal
X}_{\alpha}^{N_{f}} (\zeta_{\gamma}^{p+1})W^{p+1}_{\gamma} + {\cal
X}_{\alpha}^{N_{f}}
(\zeta_{\bar{\gamma}}^{p+1+i})W^{p+1+i}_{\bar{\gamma}})\vert_{S_{f}},
\; \; \; \mbox{for every} \; \; \alpha.
\]
${\cal X}^{S_{f}}$ is not, in general, integrable. In fact, if
$\{{\cal X}_{1}^{S_{f}}, \ldots, {\cal X}_{m}^{S_{f}}, Z_{1},
\ldots, Z_{s}\}$
is a local basis of the vector bundle $TS_{f} \to S_{f}$ then we
have that
\[
[{\cal X}_{\alpha}^{S_{f}}, {\cal X}_{\beta}^{S_{f}}] =
f^{\gamma}_{\alpha \beta} {\cal X}_{\gamma}^{S_{f}} +
\zeta^{l}_{\alpha \beta} Z_{l},
\]
for some functions $f^{\gamma}_{\alpha \beta}$ and
$\zeta^{l}_{\alpha \beta}$.

 Therefore, we must apply the integrability algorithm of
Section \ref{intalg}. Then, we obtain a sequence
$\ldots \subseteq {\cal I}_{i} \subseteq \ldots \subseteq {\cal
I}_{1} \subseteq S_{f}$,
such that ${\cal I}_{i}$ is a non-empty (closed) submanifold of
$S_{f}$, with
\beann
{\cal I}_{1} &=& \{ x \in S_{f} \vert \zeta_{\alpha \beta}^{\gamma} (x)
= 0 \}
\\
{\cal I}_{i} &=& \{ x \in {\cal I}_{i-1} \vert {\cal X}^{S_{f}}(x) \in
\Lambda^{m}\Tan_{x}{\cal I}_{i-1} \}, \; \; \; \mbox{for} \; \; i
\geq 2
\eeann
In the best cases, there exists an integer $i$ such that
${\cal I}_{i+1} = {\cal I}_{i}$. Then, ${\cal I}_{f} =
{\cal I}_{i+1} = {\cal I}_{i}$ is a submanifold of $S_{f}$ and
${\cal X}^{{\cal I}_{f}} = ({\cal X}^{S_{f}})\vert_{{\cal I}_{f}}$ is
an integrable Euler-Lagrange $m$-vector field on ${\cal I}_{f}$,
and hence it is holonomic. In fact:

\begin{teor}
If $U$ is an open subset of $M$ and $s: U \subseteq M \to {\cal
I}_{f}$ is an integral section of ${\cal X}^{{\cal I}_{f}}$ then
there exists a section $\phi: U \subseteq M \to E$ of the
projection $\pi: E \to M$ such that $s = j^{1}\phi$ and $\phi$ is
a solution to the Euler-Lagrange equations for $\Lag$.
\end{teor}
\proof
We have that
\beq \label{eq1} (\inn ({\cal
X}^{f})\Omega_{\Lag})\vert_{{\cal I}_{f}} = 0,
\eeq
\beq \label{eq2}
(\inn ({\cal X}^{f}){\cal V})\vert_{{\cal I}_{f}} = 0.
\eeq
We can assume,
without loss of generality, that $s(U) \subseteq \tilde{U}$,
with $\tilde{U}$ an open subset of $J^1\pi$ and $(x^{\alpha},
y^{A}, v^{A}_{\alpha})$ a system of local coordinates on
$\tilde{U}$. Then, since ${\cal X}^{{\cal I}_{f}}$ is locally
decomposable and $\inn ({\cal X}^{f})\omega\vert_{{\cal I}_{f}} = 1,$
we deduce that
\beq \label{*1}
{\cal X}^{{\cal I}_{f}}\vert_{\tilde{U}
\cap {\cal I}_{f}} = {\cal X}_{1}^{{\cal I}_{f}} \wedge \ldots
\wedge {\cal X}_{m}^{{\cal I}_{f}},
\eeq
 with ${\cal X}_{\alpha}^{{\cal I}_{f}} \in {\mathfrak X}(\tilde{U} \cap
{\cal I}_{f})$ given by
\beq \label{*2}
{\cal X}_{\alpha}^{{\cal I}_{f}}
= \displaystyle \left(\frac{\partial}{\partial x^{\alpha}} +
\Gamma_{\alpha}^{A} \frac{\partial}{\partial y^{A}} +
\Gamma_{\alpha \beta}^{A} \frac{\partial}{\partial
v^{A}_{\beta}}\right)\vert_{\tilde{U} \cap {\cal I}_{f}},
\eeq
for all
$\alpha$, where $\Gamma^{A}_{\alpha}$ and $\Gamma_{\alpha
\beta}^{A}$ are local functions on $\tilde{U}$.
Now, using that
\[
(\inn ({\cal X}^{{\cal I}_{f}}){\cal V})\vert_{\tilde{U} \cap {\cal
I}_{f}} = \left((\Gamma^{A}_{\alpha} - v^{A}_{\alpha}) \displaystyle
\frac{\partial}{\partial v^{A}_{\alpha}}\right)\Big\vert_{\tilde{U} \cap {\cal
I}_{f}},
\]
it follows that (see (\ref{eq2}))
\beq \label{*3}
{\cal X}_{\alpha}^{{\cal I}_{f}} = \displaystyle
\left(\frac{\partial}{\partial x^{\alpha}} + v_{\alpha}^{A}
\frac{\partial}{\partial y^{A}} + \Gamma_{\alpha \beta}^{A}
\frac{\partial}{\partial v^{A}_{\beta}}\right)\Big\vert_{\tilde{U} \cap
{\cal I}_{f}}.
\eeq
Furthermore, from (\ref{+}), we obtain that
\beq \label{contrac}
\begin{array}{lll}
\displaystyle \inn(\frac{\partial}{\partial y^{A}})
\Omega_{\Lag}&=& \displaystyle \frac{\partial^2L}{\partial
v^B_\nu\partial v^A_\alpha} \d v^B_\nu \wedge\d^{m-1}x^{\alpha}
+\left(\frac{\partial^2L}{\partial y^B\partial v^A_\alpha} -
\frac{\partial^2L}{\partial y^A\partial v^B_\alpha}\right) \d y^B\wedge
\d^{m-1}x^\alpha
  \\  & &
\displaystyle + \left(\frac{\partial^2L}{\partial y^A\partial
v^B_\alpha}v^B_\alpha
  -\frac{\partial L}{\partial y^A}+
\frac{\partial^2L}{\partial x^\alpha\partial v^A_\alpha}
\right)\d^m x
\end{array}
\eeq
 Therefore, using (\ref{eq1}), (\ref{*1}), (\ref{*2}) and
(\ref{contrac}), we conclude that 
\beq \label{+1} \displaystyle
\frac{\partial^2L}{\partial x^{\alpha} \partial v^{A}_{\alpha}} +
\frac{\partial^2L}{\partial y^{B} \partial v^{A}_{\alpha}}
v^{B}_{\alpha} + \frac{\partial^2L}{\partial v^{B}_{\nu} \partial
v^{A}_{\alpha}} \Gamma_{\alpha \nu}^{B} - \frac{\partial
L}{\partial y^{A}} = 0, \; \; \mbox{for every} \; \; A.
 \eeq
Next,
suppose that $U$ is an open subset of $M$ and that
$s: U \subseteq M \to \tilde{U} \cap {\cal I}_{f} \subseteq J^1\pi$
is an integral section of ${\cal X}^{{\cal I}_{f}}\vert_{\tilde{U}
\cap {\cal I}_{f}}$ such that the local expression of $s$ is
$s(x^{\beta}) = (x^{\beta}, s^{A}(x^{\beta}),s^{A}_{\alpha}(x^{\beta}))$.
Using (\ref{*3}) and the fact that
\[
(\Tan s)\left(\displaystyle \frac{\partial}{\partial x^{\beta}}\right) =
\displaystyle \left(\frac{\partial}{\partial x^{\beta}} +
\frac{\partial s^{A}}{\partial x^{\beta}} \frac{\partial}{\partial
y^{A}} + \frac{\partial s^{A}_{\alpha}}{\partial x^{\beta}}
\frac{\partial}{\partial v^{A}_{\beta}}\right)\Big\vert_{\tilde{U} \cap {\cal
I}_{f}},  \; \; \mbox{for every} \; \; \beta.
\]
we deduce that
\beq \label{+2} s^{A}_{\alpha} =
\displaystyle \frac{\partial s^{A}}{\partial x^{\alpha}},
\makebox[.4cm]{}  \Gamma_{\alpha \beta}^{A} \circ s =
\displaystyle \frac{\partial^2 s^A}{\partial x^{\alpha} \partial
x^{\beta}},  \; \; \mbox{for every} \; \; A,\alpha,\beta.
\eeq
 From (\ref{+2}),
it follows that there exists $\phi: U \subseteq M \to E$ a local
section of $\pi: E \to M$ such that
$s = j^1 \phi$.
Moreover, using (\ref{+1}) and (\ref{+2}), we obtain that
\[
\displaystyle \frac{\partial^2L}{\partial x^{\alpha} \partial
v^{A}_{\alpha}} + \frac{\partial^2L}{\partial y^{B} \partial
v^{A}_{\alpha}}\frac{\partial s^{B}}{\partial x^{\alpha}}+
\frac{\partial^2L}{\partial v^{B}_{\nu} \partial
v^{A}_{\alpha}}\frac{\partial^{2} s^{B}}{\partial x^{\alpha}
\partial x^{\nu}} - \frac{\partial L}{\partial y^{A}} = 0, \; \;
\mbox{for every} \; \; A.
\]
This implies that
\[
(j^i \phi)^*\left(\displaystyle \frac{\partial L}{\partial y^{A}} -
\frac{d}{dx^{\alpha}}\frac{\partial L}{\partial v^{A}_{\alpha}}\right)
= 0, \; \; \; \mbox{for every} \; \; A.
\]
In other words, $\phi$ is a solution to the Euler-Lagrange
equations associated with $\Lag$.
 \qed

\begin{remark}{\rm\label{rx}
The behaviour of the integrability algorithm in the
Lagrangian and Hamiltonian levels is the same.
Indeed, it is easy to prove that
$({\cal F}\Lag_{f})({\cal I}_{i}) = {\cal J}_i$,
 and that the map
$({\cal F}\Lag_{f})\vert_{{\cal I}_{i}}: {\cal I}_{i} \to {\cal J}_{i}$
is a diffeomorphism, for every $i$. Thus, if the integrability
algorithm in the Lagrangian level stabilizes at step $i$ then
the integrability algorithm in the Hamiltonian level also
stabilizes at step $i$ and, conversely, if the integrability
algorithm in the Hamiltonian level stabilizes at step $i$ then
the integrability algorithm in the Lagrangian level also
stabilizes at step $i$}.
\end{remark}

\section{An example: affine Lagrangian densities}
\protect\label{ex}

Consider the configuration bundle $\pi \colon E \to M$, and
$\alpha\in \Lambda_1^m \Tan^* E$. Then, $\alpha$ induces a
function $L = \hat{\alpha}\in\Cinfty(J^{1} \pi)$ as
follows: given $x \in M$ and a section $\phi\colon M\to E$, we define $L(j^1_x
\phi)$ by
$$
 L(j^1_x\phi) \eta (x)=\left[ \phi^*\alpha\right](x) \; .
$$
Note that $L(j^1_x\phi)$ is well-defined: if $\phi, \psi$ are
sections such that $j^1_x\phi = j^1_x\psi$, then $L(j^1_x\phi) =
L(j^1_x \psi)$.

Taking fibered coordinates $(x^\alpha, y^A, v^A_\alpha)$
in $J^1E$, if $\alpha = a(x^\alpha, y^A)\, d^mx +
f^\mu_B(x^\alpha, y^A)\, dy^B \wedge d^{m-1}x_\mu$, then
$$
L(x^\alpha, y^A, v^A_\alpha) = a(x^\alpha, y^A) +
f^\mu_B(x^\alpha, y^A) v^B_{\mu} \; .
$$
Thus, the Lagrangian density ${\cal L} = L \omega$ is {\sl
affine}.

A direct computation in local coordinates shows that $\Theta_{L} =
(\pi^{1*}) \alpha$ and, hence, $\Omega_{L} = (\pi)^{1*} (-
d\alpha)$. We also obtain $\widetilde{{\cal F}\Lag} = \alpha \circ
\pi^1$, and ${\cal F}\Lag=\mu\circ\alpha\circ\pi^1$. Therefore,
$\tilde{\cal{P}}=\widetilde{{\cal F}\Lag}(J^1\pi) = \alpha(E)$ is
an embedded submanifold of ${\cal M}\pi$, which is diffeomorphic
to $E$ by means of the mapping $\alpha \colon E \to
\tilde{\cal{P}} \equiv\hbox{Im}\alpha$. Since $\pi^1$ is a
surjective submersion with connected fibers, then so is
$\widetilde{{\cal F}\Lag}_0\colon J^1\pi\to{\cal P}$ 
(recall that $\widetilde{{\cal F}\Lag}_0$ is the restriction of
$\widetilde{{\cal F}\Lag}$ onto its image ${\cal P}$). Moreover,
since $\widetilde{{\cal F}\Lag}^{-1}(\widetilde{{\cal
F}\Lag})(\bar{y}) = (\pi^1)^{-1}(\pi^1(\bar{y}))$, for all
$\bar{y} \in J^1\pi$, and $\widetilde{{\cal
F}\Lag}^{-1}(\widetilde{{\cal F}\Lag})(\bar{y})\subseteq {\cal
F}\Lag^{-1}({\cal F}\Lag)(\bar{y}) \subseteq
(\pi^1)^{-1}(\pi^1(\bar{y}))$, we obtain ${\cal F}\Lag^{-1}({\cal
F}\Lag)(\bar{y}) = \widetilde{{\cal F}\Lag}^{-1}(\widetilde{{\cal
F}\Lag})(\bar{y}) = (\pi^1)^{-1}(\pi^1(\bar{y}))$, and hence the
fibers of ${\cal F}\Lag$ are connected submanifolds of $J^1\pi$.
In conclusion, affine Lagrangian systems are almost regular.

Note that the manifold ${\cal P}$ can be identified with $E$, and
the mapping ${\cal F}\Lag_0\colon J^{1}\pi \to {\cal P}$ can be
identified with the mapping $\pi^1 \colon J^{1}\pi \to E$. Hence,
the $(m+1)$-form $\Omega^0_h=(\tilde\jmath_0\circ\tilde
h)^*\Omega$ (resp. the $m$-form $\omega_0^*$) on ${\cal P}$ can be
identified with the $(m+1)$-form $ - d \alpha$ (resp.
$\pi^*(\eta)$) on $E$. Taking these identifications into account,
the constrained Hamilton equations on $E$ are
\begin{equation}\label{Hamaff}
\inn({\cal X^P})(d\alpha)=0\; ,~~~~~~\inn({\cal X^P})(\pi^*(\eta))
= 1\; .
\end{equation}
Let $\nabla^*_0$ be a connection in the bundle
$\tau_0^1\colon{\cal P}\to M$, and
 $\displaystyle{{\cal Y}^{\nabla^*_0}_\eta =
\bigwedge_{\mu=1}^{m}\left(\frac{\partial}{\partial x^\mu} +
\Gamma^A_\mu\frac{\partial}{\partial y^A}\right)}$ the
corresponding $m$-vector field on ${\cal P}$ associated with
$\nabla^*_0$ and $\eta$. A direct computation
shows that
\[
(\gamma_h)^{\nabla^*_0}_\eta = \inn({\cal
Y}^{\nabla^*_0}_\eta)\Omega_h^0 = (-1^m) \left[ \frac{\partial
f^\nu_A}{\partial x^\nu} - \frac{\partial a}{\partial y^A} +
\Gamma^B_\nu\left( \frac{\partial f^\nu_A}{\partial y^B} -
\frac{\partial f^\nu_B}{\partial y^A}\right)\right] \left( d y^A -
\Gamma^A_\mu dx^\mu\right)\; ,
\]
\[
\Omega_h^{\nabla^*_0} = \Gamma^B_\nu\left( \frac{\partial
f_B^\nu}{\partial y^A} - \frac{\partial f_A^\nu}{\partial
y^B}\right) d y^A\wedge d^m x - \frac{\partial f_B^\mu}{\partial
y^A} d y^A\wedge d y^B \wedge d^{m-1} x_\mu\; .
\]
It is easy to show that $\Omega_h^{\nabla^*_0}$ is 1-nondegenerate
if, and only if, the matrix $(\displaystyle{ f^\mu_{AB}) = \left(
\frac{\partial f^\mu_B}{\partial y^A} - \frac{\partial
f^\mu_A}{\partial y^B}\right)}$ is regular, for every $\mu \in
\{1, \dots ,m \}$. Then, the Hamiltonian constrained
system (\ref{Hamaff}) has solution. A $m$-vector field
$\displaystyle{{\cal X^P} = \bigwedge_{\mu=1}^m
\left(\frac{\partial}{\partial x^\mu} +
F^A_\mu\frac{\partial}{\partial y^A}\right)}$ is a solution to
(\ref{Hamaff}) on ${\cal P}\simeq E$ if, and only if,
$$
\left( \frac{\partial f^\mu_A}{\partial y^B} - \frac{\partial
f^\mu_B}{\partial y^A}\right)F_\mu^B = \frac{\partial a}{\partial
y^A} - \frac{\partial f^\nu_A}{\partial x^\nu}\; .
$$
Note that there are $n$ equations and $mn$ variables, and that the
rank of the matrix $\left(\displaystyle \frac{\partial
f^{\mu}_{A}}{\partial y^B} - \frac{\partial f^{\mu}_{B}}{\partial
y^A}\right)$ of type $n \times nm$ is maximum, that is, $n$. Thus, the
set of solutions of the system is an affine space of dimension
$n(m-1)$ (the solution is not unique if $m>1$).

With respect to the integrability of the solutions, a direct
computation shows that a $m$-vector field ${\cal X^P}$ solution to
(\ref{Hamaff}) is integrable if
\[
\frac{\partial F^A_\nu}{\partial x^\mu} - \frac{\partial
F^A_\mu}{\partial x^\nu} + F^B_\mu \frac{\partial
F^A_\nu}{\partial y^B} - F^B_\nu\frac{\partial F^A_\mu}{\partial
y^B} = 0, \; \; \mbox{ for all } A \mbox{ and } \mu, \nu.
\]
Otherwise, the integrability algorithm should be applied.

Taking into account the identification ${\cal P}\simeq E$,
as $\Omega_\Lag = {\cal F}\Lag_0^*(-d\alpha)$,
if ${\cal X^P}$ is a solution to the constrained Hamiltonian
equations on ${\cal P}$, then every locally decomposable $m$-vector
field ${\cal X}^{J^1\pi}$ which projects via ${\cal F}\Lag_0$ onto
${\cal X^P}$ is a solution to the equations
\[
i({\cal X}^{J^{1}\pi})(\Omega_{\Lag}) = 0, \makebox[.4cm]{}
i({\cal X}^{J^{1}\pi})\omega = 1.
\]
Let $\Psi$ be the first-order jet field with respect to the
fibration $\pi\colon E\to M$ associated to the Ehresmann
connection defined by the $m$-vector field ${\cal X^P}$. Then, the
submanifold $S$ of $J^1\pi$ where a semi-holonomic $m$ vector
field satisfying the Lagrangian equations exists is $\Psi(E)$. In
fact, if ${\cal X^S} = (\Lambda^m\Tan\Psi)\circ {\cal X^P}$ then
${\cal X}^S$ is an Euler-Lagrange $m$-vector field on $S$ for
$\Lag$, that is, ${\cal X}^S$ is a locally decomposable $m$-vector
field on $S$ and
\[
(\inn ({\cal X}^{S})\Omega_{\Lag})\vert_{S} = 0, \makebox[.3cm]{}
(\inn ({\cal X}^{S})\omega)\vert_{S} = 1, \makebox[.3cm]{} (\inn
({\cal X}^{S}){\cal V})\vert_{S} = 0.
\]
If the matrix $(f^\mu_{AB})$ is singular but there are no
higher-order constraints, the previous results remain true.
Otherwise, we will have to apply the premultisymplectic constraint
algorithm. Suppose that we have obtained the final constraint
submanifolds $N_f$ and $P_{f}$, with the submersion
$(\pi^1)\vert_{N_{f}} \colon N_{f} \to P_{f}$. Let ${\cal
X}^{P_{f}}$ be a $m$-vector field solution of the constrained
Hamiltonian equations. We have that $\pi(P_{f})$ is an open subset
of $M$ and that $\pi_{f} = \pi_{|P_{f}}: P_{f} \to \pi(P_{f})
\subseteq M$ is a fibration. Moreover, $J^1\pi_{f}$ is a
submanifold of $J^1\pi$ (see Appendix). Now, let $\Psi$ be the
first-order jet field with respect to the fibration $\pi_{f}:
P_{f} \to \pi(P_{f})$ associated to the $m$-vector field ${\cal
X}^{P_{f}}$. Then, the submanifold $S_{f}$ of $J^1\pi$ where an
Euler-Lagrange $m$-vector field for $\Lag$ along $S_{f}$ exists is
$\Psi(P_{f})$, and ${\cal X}^{S_{f}} = (\Lambda^m T\Psi)\circ
{\cal X}^{P_{f}}$ is such an Euler-Lagrange $m$-vector field (see
Theorem \ref{lmma}).

{\bf Example:} Let $\pi\colon\Real^4\to\Real^2$ be the
configuration bundle, and $L=x^2(y^1v^1_2 + y^2v^2_2) + y^1 y^2$.
In this case, $\alpha = y^1y^2dx^1\wedge dx^2 - x^2y^1dy^1\wedge
dx^1 - x^2y^2dy^2\wedge dx^1$. If $\nabla$ is the trivial
connection, $\displaystyle{{\cal Y}^\nabla_\eta =
\frac{\partial}{\partial x^1}\wedge\frac{\partial}{\partial
x^2}}$, then $\gamma^\nabla_\eta = (y^1-y^2)(dy^1-dy^2)$ and
$\Omega^\nabla_\Lag = 0$. A simple computation shows that, in this
case, $\flat_{\Omega}^{\nabla}({\bf h}) = (({\bf
h})^{H}_{\nabla},0)$. Therefore, the vector fields $Z_i$ in
Theorem 5 are all the vertical vector fields in ${\rm
V}(\bar{\pi}^1)$. Hence, the submanifold $N_1$ is characterized by
the constraint $y^1-y^2=0$. In fact, every semi-holonomic
$2$-vector field in $N_1$ is an Euler-Lagrange $2$-vector field
for this problem.

\section*{Appendix: $m$-vector fields and Ehresmann connections in fibre bundles}
\protect\label{mvfec}

(See \cite{EMR-98,LMM-96,LMMa-2002} for the proofs and other details
 about the results in this section).

Let $F$ be a $N$-dimensional differentiable manifold.
Sections of $\Lambda^m(\Tan F)$ are called
{\sl multivector fields} in $F$, or more precisely,
$m$-{\sl vector fields} in $F$
(they are contravariant skew-symmetric tensors of order $m$ in $F$).
The space of $m$-vector fields is denoted by ${\cal V}^{m}(F)$.
 ${\cal X}\in{\cal V}^m(F)$ is {\sl locally decomposable} if,
for every $p\in F$, there exists an open neighbourhood $U_p\subset F$
and $Y_1,\ldots ,Y_m\in\vf (U_p)$ such that
${\cal X}\feble{U_p}Y_1\wedge\ldots\wedge Y_m$.
 We denote by $\vf^m (F)$ the set of locally decomposable 
$m$-vector fields in $F$.
 Contraction of $m$-vector fields and tensor fields in $F$
 is the usual one.

We can define an equivalence relation:
if ${\cal X},{\cal X}'\in\vf^m(F)$ are non-vanishing $m$-vector fields,
and $U\subseteq F$ is a connected open set,
then ${\cal X}\stackrel{U}{\sim}{\cal X}'$ if there exists a
non-vanishing function $f\in\Cinfty (U)$ such that
${\cal X}'\feble{U}f{\cal X}$. Equivalence classes
are denoted by $\{ {\cal X}\}_U$.
There is a one-to-one correspondence between the set of $m$-dimensional
orientable distributions $D$ in $F$ and the set of the
equivalence classes $\{ {\cal X}\}_F$ of non-vanishing, locally decomposable
$m$-vector fields in $F$.
If ${\cal X}\in\vf^m(F)$ is non-vanishing and locally decomposable,
 the distribution associated
with the class $\{ {\cal X}\}_U$ is denoted ${\cal D}_U({\cal X})$
(If $U=F$ we write ${\cal D}({\cal X})$).
A non-vanishing, locally decomposable
$m$-vector field ${\cal X}\in\vf^m(F)$ is said to be {\sl integrable}
if its associated distribution ${\cal D}_U({\cal X})$ is integrable.
Of course, if ${\cal X}\in\vf^m(F)$ is integrable,
then so is every $m$-vector field in its equivalence class $\{ {\cal X}\}$,
and all of them have the same integral manifolds.
Moreover, from {\sl Frobenius' theorem},
a non-vanishing and locally decomposable $m$-vector field is integrable
 if, and only if, ${\cal D}({\cal X})$ is involutive.

Now, let $\kappa\colon F\to M$ be a fibre bundle ($\dim\, M=m$).
We are concerned with the case where the integral manifolds of
integrable $m$-vector fields in $F$ are sections of $\kappa$.
Thus, ${\cal X}\in\vf^m(F)$ is said to be {\sl $\kappa$-transverse}
if, at every point $y\in F$,
$(\inn ({\cal X})(\kappa^*\eta))_y\not= 0$, for every $\eta\in\df^m(M)$
such that $\eta (\kappa(y))\not= 0$.
Then, if ${\cal X}\in\vf^m(F)$ is integrable, it is $\kappa$-transverse
 if, and only if,
its integral manifolds are local sections of $\kappa\colon F\to M$.
In this case, if $\phi\colon U\subset M\to F$
is a local section with $\phi (x)=y$ and $\phi (U)$ is
the integral manifold of ${\cal X}$ through $y$,
then $\Tan_y({\rm Im}\,\phi)$ is ${\cal D}_y({\cal X})$.
Integral sections $\phi$ of ${\cal X}$ can be characterized
by the condition
$$
\Lambda^m\Tan\phi=f{\cal X}\circ\phi\circ\varrho_M
$$
where
$\Lambda^m\Tan\phi\colon\Lambda^m\Tan M\to\Lambda^m\Tan F$
is the natural lifting of $\phi$,
$\varrho_M\colon\Lambda^m\Tan M\to M$
is the natural projection,
and$f\in\Cinfty (F)$ is a non-vanishing function
(observe that this characterizes the entire class
$\{ {\cal X}\}$ of integrable $m$-vector fields).

Let $\nabla$ be an Ehresmann connection in the fibration
$\kappa\colon F \to M$. As is known, it defines  a horizontal
subbundle ${\rm H}(\nabla) \subset \Tan F$, such that $\Tan F = {\rm
H}(\nabla) \oplus {\rm V}(\kappa)$, where ${\rm V}(\kappa)$ is the
$\kappa$-vertical subbundle. If $y \in F$, then
${\rm H}_{y}(\nabla)= Im \nabla(y)$. Thus,
we have the horizontal distribution associated
with the connection $\nabla$.
The connection $\nabla$ is said to be flat (respectively, orientable) if the
horizontal distribution is completely integrable (respectively,
orientable).

Classes of locally decomposable and
$\kappa$-transverse $m$-vector fields $\{{\cal X}\} \subseteq \vf^m (F)$
are in one-to-one correspondence with orientable Ehresmann connections
 $\nabla$ in $\kappa \colon F \to M$. This correspondence is
given by the fact that the horizontal subbundle associated
with $\nabla$ is ${\cal D}({\cal X})$. Thus,
classes of integrable locally decomposable and $\kappa$-transverse
$m$-vector fields correspond to flat orientable Ehresmann
connections.

A connection $\nabla$ in the fibration $\kappa \colon F \to M$
induces a splitting $\Tan^{*}F={\rm H}^*(\nabla)\oplus{\rm V}^*(\kappa)$, where
\[
{\rm H}^*_{y}(\nabla) = {\rm V}_{y}(\kappa)^0, \quad
{\rm V}^*_{y}(\kappa) = {\rm H}_{y}(\nabla)^0.
\]
Here, ${\rm V}_{y}(\kappa)^0 \subset \Tan_{y}^{*}F$
(respectively, ${\rm H}_{y}(\nabla)^0 \subset \Tan_{y}^{*}F$) denotes the
annihilator of the subspace ${\rm V}_{y}(\kappa) \subset \Tan_{y}F$
(respectively, ${\rm H}_{y}(\nabla) \subset \Tan_{y}F$). The
splittings $\Tan F = {\rm H}(\nabla) \oplus {\rm V}(\kappa)$
and $\Tan^{*}F = {\rm H}^*(\nabla)\oplus{\rm V}^*(\kappa)$
may be extended to the tensor bundles
\beq
\label{ae1}
\Lambda^l \Tan F=\bigoplus_{r,s=0,\ldots ,l;\ r+s=l}
(\Lambda^r{\rm H}(\nabla)\oplus\Lambda^s{\rm V}(\kappa))
\eeq
\beq
\label{ae1'}
\Lambda^k \Tan^*F=\bigoplus_{p,q=0,\ldots ,k;\ p+q=k}
(\Lambda^p{\rm H}^*(\nabla)\oplus\Lambda^q{\rm V}(\kappa)^*)
\eeq
Thus, for every $X \in \vf (F)$, we obtain that $\inn(X) \nabla
\equiv X^{H}_{\nabla}$ is an horizontal vector field, that is, a
section of ${\rm H}(\nabla) \to F$. $X^{H}_{\nabla}$ is the
horizontal component of $X$, and we write $X = X^{H}_{\nabla} +
X^{V}_{\nabla}$, where $X^{V}_{\nabla} = X - X^{H}_{\nabla}$ is a
$\kappa$-vertical vector field. Moreover, if $\alpha \in \df^{1}(F)$,
then we have that $\inn (\alpha)\nabla \equiv \alpha_{\nabla}^{H} \in
\df^{1}(F)$ is an horizontal $1$-form, that is, a section of ${\rm
H}(\nabla)^* \to F$. $\alpha^{H}_{\nabla}$ is the horizontal component of
$\alpha$, and we write $\alpha = \alpha^{H}_{\nabla} +
\alpha^{V}_{\nabla}$, where $\alpha^{V}_{\nabla} = \alpha -
\alpha^{H}_{\nabla}$ is a $\kappa$-vertical $1$-form with respect to
the connection $\nabla$, that is, it vanishes under the action of
every horizontal vector field associated with the connection
$\nabla$. Furthermore, if $X \in \vf (F)$ is a $\kappa$-vertical
vector field, then $\inn(X)\alpha^{H}_{\nabla} = 0$. In addition, if
${\cal X} \in\vf^k(F)$ and $\beta \in \df^l(F)$, the
splittings (\ref{ae1}) and (\ref{ae1'}) allow us to make the
following decomposition
\[
{\cal X} = \sum_{r,s = 0; r+s=k} {\cal X}^{(r,s)}_{\nabla}, \quad \beta =
\sum_{p,q = 0; p+q = l}\beta^{(p,q)}_{\nabla},
\]
where the superscripts $(i, j)$ denote the horizontal and vertical
parts respectively, of the $k$-vector field ${\cal X}$ and the
$l$-form $\beta$.

Finally, if $\nabla$ is an Ehresmann connection in the fibration
$\kappa \colon F \to M$ and $y\in F$ then the map
\[
\Lambda^{k}\Tan_y\kappa_{{\rm H}(\nabla)}\colon \Lambda^{k}{\rm H}_{y}(\nabla)
\to\Lambda^{k}\Tan_{\kappa(y)}M,
 \quad 1 \leq k \leq dim\, M = m,
\]
is a linear isomorphism and the inverse morphism
$(\Lambda^m\Tan_y\kappa_{{\rm H}(\nabla)})^{-1}\colon \Lambda^m\Tan_{\kappa(y)}M
\to \Lambda^m{\rm H}_{y}(\nabla)$
is just the horizontal lift at $y$ induced by $\nabla$.
Denoting by $\Lambda^m(\kappa_{{\rm H}(\nabla)})_*^{-1}$
the natural extension of this map to $m$-vector fields on $M$,
one may consider $\Lambda^m(\kappa_{{\rm H}(\nabla)})^{-1}_*({\cal X})$,
the horizontal lift of ${\cal X}\in\vf^m(M)$,
as the $m$-vector field on $F$ given by
\[
[\Lambda^m(\kappa_{{\rm H}(\nabla)})^{-1}_*({\cal X})](y) =
(\Lambda^m\Tan_y\kappa_{{\rm H}(\nabla)})^{-1}({\cal X}(\kappa(y))),
 \quad \mbox{for every} \ y\in F.
\]
In particular, if ${\cal X}_{\eta}$ is the $m$-vector field on $M$ characterized
by the condition
\[
\alpha_{1} \wedge \dots \wedge \alpha_{m} =
{\cal X}_{\eta}(\alpha_{1}, \dots , \alpha_{m})\eta
\quad , \quad
\mbox{\rm for every $\alpha_{1}, \dots , \alpha_{m} \in \df^{1}(M)$}
\]
one may define the $m$-vector field
${\cal Y}_{\eta}^{\nabla}\in\vf^m(F)$ by
${\cal Y}_{\eta}^{\nabla} =
\Lambda^m(\kappa_{{\rm H}(\nabla)})^{-1}_*({\cal X}_{\eta})$.
 Note that ${\cal Y}_{\eta}^{\nabla}$ is 
a locally decomposable and $\kappa$-transverse
$m$-vector field on $F$, verifying that
$\inn({\cal Y}_{\eta}^{\nabla})\omega=1 $, and that the distribution
${\cal D}({\cal Y}_{\eta}^{\nabla})$ 
is just the horizontal distribution associated
with the connection $\nabla$.

If $C$ is a submanifold of $F$, and
${\cal X}_{C}$ is a locally decomposable $m$-vector field on $C$
such that
\[
\inn({\cal X}_{C}(y))\omega(y) = 1, \quad \mbox{for every} \; \; y \in C
\]
then $\kappa\vert_C\equiv\kappa_C\colon C \to M$ is a submersion.
In fact, if $y \in C$ and
${\cal X}_C(y)=X_C^1(y)\wedge\dots\wedge X_C^m(y)$,
with $X_C^i(y)\in\Tan_yC$, then
$$
\eta(\kappa(y))(\Tan_y\kappa_C(X_C^1(y)),\dots ,\Tan_y\kappa_C(X_C^m(y)))=1
$$
 This implies that $\{\Tan_y\kappa_C(X_C^1(y)),\dots ,\Tan_y\kappa_C(X_C^m(y))\}$
is a basis of $\Tan_{\kappa(y)}M$, and thus, $\Tan_y\kappa_C\colon
\Tan_yC\to\Tan_{\kappa(y)}M$ is an epimorphism. Therefore,
$\kappa(C)$ is an open subset of $M$ and $\kappa_C\colon C
\to\kappa(C)$ is a fibre bundle. Consequently, ${\cal X}_{C}$
defines an oriented Ehresmann connection in the fibration
$\kappa_C\colon C\to \kappa(C)$ which, in the terminology of
\cite{LMM-96,LMMa-2002}, is said to be an (oriented) {\sl
Ehresmann connection in the fibration $\kappa \colon F \to M$
along the submanifold $C$}. Note that the canonical inclusion
$\iota: J^{1}\kappa_{C} \to J^1\kappa $ is an embedding and, thus,
$J^1\kappa_{C}$ is a submanifold of $J^1\kappa$.

\begin{remark}{\rm\label{r3}
It is well-known
\cite{Sa-89} that there exists a one-to-one correspondence between
Ehresmann connections in the fibration $\kappa \colon F \to M$ and
first-order jet fields with respect to $\kappa$, that is, sections of
the fibration $\kappa^1\colon J^1F\to F$.
In fact, let $\nabla$ be a connection in the fibration $\kappa\colon F\to M$,
(that is, an element of
$\Gamma (E,\kappa^*\Tan^*M)\otimes\Gamma (F,\Tan F)$),
such that $\nabla^*\alpha =\alpha$, for every
$\kappa$-semibasic form $\alpha\in\df^1(F)$),
and ${\rm H}(\nabla)$ the associated horizontal subbundle.
If $(\Tan_y\kappa_{{\rm H}(\nabla)})^{-1}$ denotes the horizontal lift at $y$;
for every $y\in F$, let $\phi\colon M\to F$ be a section of $\kappa$
passing through $y$, such that
$$
\Tan_{\kappa(y)}\phi=\Tan_y\kappa_{{\rm H}(\nabla)})^{-1}\colon
\Tan_{\kappa(y)}M\to{\rm H}_y(\nabla)\subset \Tan_{y}F
$$
then we define the map
$$
\begin{array}{ccccc}
\psi^{\nabla}&\colon& F & \to & J^1F \\
& & y & \mapsto & (j^1\phi)(\kappa(y))
\end{array}
$$
which is a section of the fibration $\kappa^1\colon J^1F \to F$. Conversely,
given a section $\psi^{\nabla}\colon F\to J^1F$, for every $\bar y\in J^1F$
with $\kappa^1(\bar y)=y$,
and a representative $\phi\colon M\to F$ of $\bar y$,
we define the horizontal subspace ${\rm H}_y(\nabla):={\rm Im}\Tan_y\phi$,
and ${\rm H}(\nabla):=\cup_y{\rm H}_y(\nabla)$.
Thus we have identified the fibre $J^1_yF=(\kappa^1)^{-1}(y)$
with the set
$$
\{ {\bf h}_{y}\in
\Tan_{\kappa(y)}^*M \otimes \Tan_{y}F \ \mid \ \Tan_y\kappa\circ{\bf h}_{y}=Id\}
$$

In particular, if we have a connection or, what is equivalent, a
class of $\kappa$-transverse, locally decomposable $m$-vector
fields in the fibration $\kappa \colon F \to M$, along a
submanifold $C$ of $F$, and a representative ${\cal X}_{C}$ of
this class, then $\kappa(C)$ is an open subset of $M$, $\kappa_C =
\kappa\vert_C\colon C \to \kappa(C)$ is a fibration, and ${\cal
X}_{C}$ may be seen as a section $\psi^\nabla_{C}$ of the
fibration $\kappa_C^1\colon J^1\kappa_C\to C$. Thus,
$\psi^\nabla_{C}(y)$ is identified with a linear map from
$\Tan_{\kappa(y)}M$ onto $\Tan_{y}C$, that is, an element ${\bf
h}_{y}\in\Tan_{\kappa(y)}^{*}M\otimes \Tan_{y}C$, and
\[
(\Tan_y\kappa_C\circ\psi_C)(y) =
(\Tan_y\kappa_C\vert_{\Tan_yC}\circ\psi^\nabla_C)(y) = Id, \quad
\mbox{for every} \ y \in C.
\]
}\end{remark}

\subsection*{Acknowledgments}

We acknowledge the financial support of 
{\sl Ministerio de Educaci\'on y Ciencia},
projects BFM2002-03493, BFM2003-01319 and MTM2004-7832.
We thank Mr. Jeff Palmer for his
assistance in preparing the English version of the manuscript.

 \end{document}